\documentclass{mn2e}
\usepackage{epsfig}
\usepackage{amsmath}

\title[Line branching]{Mass-loss
rates for hot luminous stars: the influence of line branching}
\author[S. A. Sim]{S. A. Sim\thanks{s.sim@imperial.ac.uk} \\
Astrophysics Group, Imperial College London,
Blackett Laboratory, Prince Consort Road, London, SW7 2BW, UK}

\date{\today}

\pagerange{\pageref{firstpage}--\pageref{lastpage}}
\pubyear{2003}

\begin{document}
\maketitle
\label{firstpage}

\begin{abstract}
The effect of photon frequency
redistribution by line branching on mass-loss rates for hot luminous stars 
is investigated.
Monte Carlo simulations are carried out for a range of OB-star models
which show that previous mass-loss calculations which neglect 
non-resonance line scattering overestimate mass-loss rates 
for luminous O stars by $\sim$~20 per cent. For 
luminous B stars the effect is
somewhat larger, typically $\sim$~50 per cent.
A Wolf-Rayet 
star model is used to investigate line branching in the strong wind limit. 
In this case the effect of line branching is much greater, 
giving mass-loss rates that are smaller by a factor $\sim 3$ from 
computations which neglect branching.
\end{abstract}

\begin{keywords} 
radiative transfer --
methods: numerical --
stars: mass-loss --
stars: individual: $\zeta$~Pup --
stars: Wolf-Rayet
\end{keywords}

\section{Introduction}

It has been known for over thirty years that luminous, hot stars have powerful
stellar winds (Morton 1967). 
These winds are important in a variety of astrophysical contexts, most 
notably because of their dynamical influence on the interstellar medium,
their effect on stellar evolution and their role in metal enrichment of
galaxies.

The proposal that these winds are driven by
radiation pressure, in particular by photon scattering in spectral lines,
was developed by Lucy \& Solomon (1970) and is now generally accepted
(Kudritzki \& Puls 2000).
Since the identification of the mass-loss mechanism, 
attention has been concentrated on the 
development of techniques which can predict reliable 
mass-loss rates. Currently, the most accurate mass-loss rates for hot 
stars are obtained from Monte Carlo
simulations of radiation transport in winds (de Koter, Heap \& Hubeny 1997;
Vink, de Koter \& Lamers 1999,
2000; hereafter VdKL99, VdKL00 respectively), 
an approach to mass-loss calculations first applied by Abbott \& Lucy 
(1985, hereafter AL85). 
In this method, the global constraint of energy conservation is
used to determine the
mass-loss rate algebraically while the wind terminal velocity is
fixed at the observed value. This is in
contrast to other techniques, such as `CAK' theory 
(Castor, Abbott \& Klein 1975),
where the mass-loss rate is an eigenvalue determined at a critical point and
the terminal velocity is calculated.
Monte Carlo simulations have also been used to study mass-loss as
a function of metallicity (Vink, de Koter \& Lamers 2001) and
mass-loss from luminous blue variables (Vink \& de Koter 2002).

A weakness of previous determinations of mass-loss rates from Monte 
Carlo
simulations is the assumption that all line scattering events may be
treated as resonance scatterings. This assumption was made by AL85 
and has been followed by others in subsequent work (including 
the mass-loss calculations of
de Koter et al. 1997 and VdKL99,00). It has the advantage of 
significantly simplifying the
computation and can be justified for many of the strong lines, even some
non-resonance lines, in several important ions (see AL85). 
However, although remarkably successful for computing
mass-loss rates, the assumption 
is not valid in general: Lucy (1999b) included 
downwards branching in a Monte Carlo spectral 
synthesis
code for supernovae and found this to have a significant impact on the
emergent spectrum.

Following recent developments of the Monte Carlo 
energy-packet technique (Lucy 2002,
2003), it has become feasible to 
lift
the resonance scattering assumption and allow branching to occur
in line scattering events. Branching may effect mass-loss rates since it
allows photon leakage from one part of the spectrum to another
(see e.g. the discussion by 
Owocki \& Gayley 1999). If only
resonance scattering is permitted, 
the frequency of
a given photon 
packet
is gradually redshifted in the frame co-moving with the 
stellar wind (which accelerates outwards in the star's reference frame):
photons have no opportunity to jump in frequency.
Branching alters this by allowing photons to be re-emitted
in a different part of the spectrum 
following a line scattering event. The net effect of 
line branching on many photons is to preferentially redistribute photons 
from regions of the spectrum with many strong spectral lines 
(e.g. ultraviolet) to relatively sparse
regions (e.g. infrared). 
This will reduce the net line force since photons undergo fewer
line scatterings.
The purpose of this paper is to quantify this effect and determine
the extent by which the assumption of resonance scattering
causes mass-loss rates to be overestimated.
It is to be expected that the magnitude of the effect will
depend on the typical number of
scatterings that photons undergo as they traverse the wind:
in the limit where no photons are scattered more than once
(single scattering limit) the frequency of re-emission is irrelevant to the
mass-loss calculation meaning that line branching can have no effect.

Most of the results presented in this paper relate to OB stars, the stars
for which VdKL00 have computed mass-loss rates. However, 
calculations appropriate for Wolf-Rayet (W-R) stars are also presented.
The winds of W-R stars present the greatest challenges to the theory of
radiation driven mass-loss, requiring significantly higher efficiency 
in the transfer of momentum from the radiation field to the matter than
OB star winds. 
Lucy \& Abbott (1993, hereafter LA93) 
investigated line driving in W-R stars and concluded
that it was possible to obtain wind performance numbers $\eta
\equiv \Phi v_{\infty} c / L_{*}$ which are significantly in excess of one,
as required for W-R winds
($\Phi$ is the mass-loss rate, $v_{\infty}$ is the terminal velocity of the
wind and $L_{*}$ is the luminosity of the wind-free star).
de Koter et al. (1997) performed
Monte Carlo simulations with relatively 
sophisticated models for specific WN5h 
and Of/WN
stars (as classified by 
Crowther \& Dessart 1998) to
compute line-driven 
mass-loss rates for comparison 
with values determined 
from H$\alpha$ observations. 
They found that the computed mass-loss rates were 
too small by factors $< \sim 2$ (note,
however, that
there are significant uncertainties in the mass-loss rates determined
from observations).
A complete understanding of the mass-loss mechanism in W-R stars does not yet 
exist, but it seems certain that line and continuum driving have some role
(see e.g. Nugis \& Lamers 2002;
Lamers \& Nugis 2002). This investigation
is of particular relevance to W-R stars because the high density and ionisation
stratification of W-R winds means that multiple scattering is much more
prevalent than in OB stars. Since the effect of line branching on the 
radiative acceleration can only be important when multiple
scattering occurs, W-R stars should exhibit the
most extreme effects.

Subsequent to the commencement of the work presented here, Onifer \&
Gayley (2003) have published a complementary study of frequency
redistribution in stellar winds. They present a
parameterised analytical theory 
appropriate for an idealised spectrum consisting of two frequency
domains, one line-rich and the other line-poor. However, they did not
perform any calculations with a real line list meaning they were unable to
obtain quantitative estimates of the effect of frequency redistribution
on mass-loss rates. In contrast, the numerical computations that are
presented here do not require any idealisations of the distribution of
line opacity across the spectrum and use a real line list, allowing
quantitative results to be derived.

Study of line branching is also relevant for predictions of
ionising fluxes from hot stars. Calculations which neglect
line branching, for example {\it CoStar} models (Leitherer et al. 1996,
Schaerer \& de Koter 1997), may overestimate the ionising flux as a
result (Crowther et al. 1999). The new Monte Carlo techniques employed
here (Lucy 2002,2003) will readily allow this effect to be quantified in the 
near future.

In Section 2, the stellar wind models that will be used for the mass-loss
calculations are discussed. The Monte Carlo approach used for computing
mass-loss rates is described in Section 3, and Section 4 introduces the
atomic data used in the calculations. The results for models of OB stars
are presented in Section 5 and for a W-R star model in Section 6. Conclusions
are drawn in Section 7.

\section{Models}

\subsection{Velocity and density}

As discussed above the objective of this paper is a {\it differential} study
of the effects of line branching and not the development of high quality
stellar wind models. Therefore, simple ``core-halo'' models are adopted for 
the sake of tractability. They have been constructed following AL85
and LA93.

The one-dimensional velocity law 

\begin{equation}
v(r) = v_c + (v_{\infty} - v_c) \left( {1 - \frac{R_{c}}{r}} \right)^{\beta}
\end{equation}
is adopted where $R_{c}$ is the core radius (the base of the wind)
and $v_c$ is the wind velocity at $R_{c}$.
The adopted values of $R_c$, $v_c$, $v_{\infty}$ and $\beta$ are given
in Sections 5 and 6. The mass density $\rho$ is determined from the equation of
mass conservation

\begin{equation}
4 \pi r^2 \rho v = \Phi \; \; .
\end{equation}

\subsection{Ionisation and excitation}

The approach to ionisation and excitation recommended by Lucy (2003) is
adopted here. Namely, only approximate non-LTE solutions are pursued and 
analytic formulae are used for both ionisation and excitation. This 
approach has been used in many previous studies (including AL85 and, 
in part, VdKL99,00) and 
is particularly suitable for this study since it decouples the Monte Carlo
simulation of the radiation field from the computation of level populations. 
This provides two advantages: first the need for iteration when 
constructing models is removed and, 
more importantly, it allows direct study of the effect of 
line branching on the radiation field in isolation, without any
feedback influence
on the level populations.

For ionisation, the modified nebular approximation is adopted

\begin{equation}
\frac{N_{j+1}N_{e}}{N_{j}} = \left[{\zeta_j W + (1-\zeta_j)W^2}\right]
\sqrt{\frac{T_e}{T_R}} \left({\frac{N_{j+1}N_{e}}{N_{j}}}\right)^{\mbox{*}}_{T_{R}}
\end{equation}
where $N_{j}$ is the number density of the $j$th ionisation stage, $N_{e}$ is
the free electron number density, $W$ is the dilution factor which is
defined in terms of the photospheric radius ($R_{p}$) following LA93,
$\zeta_j$ is the fraction of recombinations to ion $j$
that go direct to the ground state, 
$T_{e}$ is the electron temperature
and $T_{R}$ is the radiation temperature.
The last bracket in equation~(3)
is the ionisation fraction computed using the Saha equation
at temperature
$T_R$. 
The partition functions used for this calculation are non-LTE, having
been computed using diluted Boltzmann population
ratios for non-metastable levels (see below).
The $\zeta$-values used by
AL85 have been adopted throughout this paper. 
The radiation and electron temperatures used for the 
calculations
are discussed in Sections 5 and 6.

When determining level populations, following AL85 and
VdKL99, a distinction is made between metastable levels and those
with permitted electric dipole decay routes. For a metastable level $i$, the
population relative to the ground state $1$ is assumed to be given by the
Boltzmann formula

\begin{equation}
\frac{n_i}{n_1} = \left({\frac{n_{i}}{n_{1}}}\right)^{\mbox{*}}_{T_{R}}
\end{equation}
while for all other levels it is assumed that the population is driven 
by radiation such that

\begin{equation}
\frac{n_i}{n_1} = W \left({\frac{n_{i}}{n_{1}}}\right)^{\mbox{*}}_{T_{R}}\;.
\end{equation}

Equations (3), (4) and (5) allow all level populations to be determined for a given
value of $\rho$. Although approximate, Springmann \& Puls (1998) have 
demonstrated that in a typical O star, these formulae provide excellent
approximations to ion/level populations obtained from much more detailed
non-LTE calculations.

\section{The radiation field}

The radiation field is simulated by a Monte Carlo calculation using the 
Macro Atom approach developed by Lucy (2002). The principles of the
method are discussed by Lucy (2002,2003). The details of the Monte Carlo
simulation used here are presented below, with particular emphasis on the
differences between the calculations discussed by Lucy and those performed
here.

\subsection{Discretization of model}

For the Monte Carlo calculation, the wind models (Section 2) are
discretized into a series of thin spherical shells. The velocity is allowed
to vary in accordance with equation~(1) throughout the model, but all other
parameters are taken to be constant within each shell. 
To evaluate the density ($\rho$), for each shell 
equation (2) is used for twenty different
values of $r$ which are equally spaced in $x=R_{c}/r$ 
between the inner and outer boundaries of the shell.
The adopted density for the shell is the $x^{-4}$-weighted mean of these twenty
values. (This weighting is chosen to preserve mass in spherical geometry.)
The dilution factor ($W$) is evaluated at the mid-point of each shell.
The innermost shell
lies on the surface of the core $r=R_c$, the outermost shell extends to 
$r=100R_{c}$. Typically 100 shells of equal 
thickness in $x=R_{c}/r$ are used in the calculations.

\subsection{Creation of {\boldmath $r$}-packets}

The Monte Carlo quanta used in the simulation are indivisible packets of
radiative energy, termed $r$-packets by Lucy (2002). The Monte Carlo method
imposes radiative equilibrium by conserving both the number of packets
and their co-moving energy throughout the calculation. Thus packets are not
created in the domain of the simulation but are only launched into the inner
shell (from the surface of the star) and propagate until they 
either re-enter the star through the lower boundary of the innermost shell
or emerge from the upper boundary of the outermost shell.

The radiation field incident from the stellar core is modelled by a 
black-body spectrum at the temperature of the stellar surface $T_e(R_c)$.
A lower limit to the wavelength of the incident spectrum is imposed at
$\lambda_{\mbox{\scriptsize min}} = 228$~\AA~ on the assumption that
higher energy photons are immediately removed via bound-free absorption by
He~{\sc ii}.

The frequencies $\nu$ of the incident $r$-packets are equally spaced in 
$\ln \nu$, and their energies are assigned according to the
Planck spectrum. Packets are created with wavelengths up to 
$100 \lambda_{\mbox{\scriptsize min}}$.
The initial direction cosines $\mu$ are chosen randomly, assuming no 
limb-darkening. Typically, $2 \times 10^6$ packets are used
in the calculation.

\subsection{Propagation of packets}

As the packets propagate through the wind they undergo interactions with
the material in the wind. Immediately after their creation at the lower 
boundary, after a packet crosses a shell boundary  
and also after any interaction with wind material, the optical depth
that a packet will traverse before its next interaction is chosen at 
random using $\tau = - \ln z$ where $z$ is a random number in the interval
0 to 1. Contributions to the optical depth along the packet's flight path
from both continuous and line opacity are then considered to determine 
whether the packet undergoes a continuum scattering, a line absorption or
whether the packet will cross a boundary into another shell. The method
for selecting which event occurs follows closely that described by 
Lucy (2003), sec. 5, and the reader is referred to that paper for a 
full explanation. The subsections that follow briefly explain how
the continuum and line contributions to the opacity are computed.

\subsubsection{Electron scattering}

The only continuum process included in the calculations is scattering by
free electrons. Although bound-free processes are important in spectral
synthesis, they are not the dominant contribution to the radiation driving
force in OB stars and are therefore not important in this differential study
of line branching.

Electron scattering is assumed to be isotropic and coherent in the 
co-moving frame.

\subsubsection{Line absorption}

The large velocity gradients in a stellar wind mean that line processes
can be treated in the Sobolev limit. Lucy (2002) gives expressions for the
Sobolev radiative rates and 
optical depth for the case of an atmosphere in homologous expansion
(his equations 20, 21 and 22). However, for the case of non-homologous
expansion in a stellar wind, these expressions must be generalised.
The radiative rates for spontaneous emission and absorption in the
transition between upper level $i$ and lower level $j$ are given by
the standard expressions

\begin{equation}
R_{ij} = A_{ij} \beta_{ij} n_{i} \; \; \; \; \mbox{and} \; \; \; \; 
R_{ji}=(B_{ji}n_{j}-B_{ij}n_{i})\overline{J}_{ji}
\end{equation}
where $A$ and $B$ are the usual Einstein coefficients, 
the $n$'s are level 
populations (number densities) and $\overline{J}_{ji}$ is the integrated 
mean intensity in the line.
The escape probability, 
$\beta_{ji}$ is given by (Klein \& Castor 1978)

\begin{equation}
\beta_{ji} = \frac{1}{2}\int_{-1}^{1} \frac{1 - e^{-\tau_{s}(\mu)}}{\tau_{s}(\mu)} \; \mbox{d}{\mu}
\end{equation}
and the Sobolev optical depth $\tau_{s}(\mu)$ is

\begin{equation}
\tau_{s} = - (B_{ji} n_{j} - B_{ij} n_{i})\frac{h \nu_{ij}}{4\pi} \frac{1}
{\mbox{d}\nu/\mbox{d}s}
\end{equation}
where $\nu_{ij}$ is the line frequency and 

\begin{equation}
\frac{\mbox{d}\nu}{\mbox{d}s}=-\frac{\nu_{R}}{c}
\left[{(1-\mu^2)\frac{v}{r} + \mu^2 \frac{\mbox{d}v}{\mbox{d}r}}\right]
\end{equation}
is the rate of change of co-moving frequency with path-length for a 
photon with rest frame frequency $\nu_{R}$
at radius $r$ travelling in the direction specified by $\mu$.
The dependence of $\tau_{s}$ on ${\mbox{d}\nu}/{\mbox{d}s}$ does not
significantly complicate the Monte Carlo calculation since 
${\mbox{d}\nu}/{\mbox{d}s}$ may be taken as constant along any
given photon trajectory within a shell,
provided that the shells are sufficiently thin.

In this formulation, the interaction between a packet and a spectral
line with which is comes into resonance is described by a single scattering
event. In reality, when a photon enters resonance with a line it may scatter
many times in that line before it is redshifted out of the region
in which interaction occurs (the {\it resonance zone}). Since 
thermalisation is not included here, these multiple interactions with a 
particular line are readily described by a single scattering event and the
usual
Sobolev escape probability (see de Koter et al. 1997 for a discussion
of thermalisation during multiple interactions within a resonance zone).
Accordingly, throughout this paper the term {\it single scattering} is
used to refer to photon packets that interact with {\it one line} during 
the Monte Carlo simulation, even though this
interaction may physically consist of several scatterings. {\it Multiple
scattering} refers to energy packets that undergo two or more independent 
interactions with {\it different} spectral lines during their flight through 
the model.

The Sobolev optical depth of a line is used to determine whether or not
a packet is absorbed when it comes into resonance with the line. If absorption
occurs the Macro Atom method developed by Lucy (2002) is used to determine
the atomic level from which re-emission occurs and a random number is
used to determine in which of the transitions from the level the packet is
emitted, following Lucy (2003). The Macro Atom formalism does allow for
collisional, bound-free and free-free processes but, for simplicity, these
are not included here.

The calculation of the Macro Atom upward jumping probabilities
requires $\overline{J}$ to be known for each spectral line. 
Lucy (1999b,2003) has shown how efficient Monte
Carlo estimators may be constructed for integrals of the radiation
field. 
Following Lucy, the estimator used here 
for a transition between upper level $i$ and lower level $j$ in a 
particular shell $p$ is

\begin{equation}
\overline{J}_{ji} = - \frac{1}{4\pi\Delta t}\frac{1}{V_p}
\sum \frac{1}{\mbox{d}\nu/\mbox{d}s}
\frac{1}{\tau_{s}}(1 - e^{-\tau_{s}}) \epsilon
\end{equation}
where $\Delta t$ is the time interval represented by the Monte
Carlo simulation and $V_p$ is
the volume of the shell $p$. The sum runs over all packets which come
into resonance with the line in shell $p$ during the
simulation. $\epsilon$ is the packet energy in the
co-moving frame and $\tau_{s}$ is the 
Sobolev optical depth of the transition. Note that
since $\tau_{s}$ is angle dependent every packet encounters a different optical
depth: thus the factors involving $\tau_{s}$ 
must be placed inside the summation.
For the branching calculations presented in the later sections of this paper,
the $\overline{J}_{ji}$-values are calculated iteratively for 
each model, with initial 
values taken as a dilute black-body radiation field. The values of 
$\overline{J}_{ji}$ 
were found to converge to sufficient accuracy for the mass-loss
calculations within only one or two iterations.

For all the models discussed in this paper, in addition to finding mass-loss
rates when branching is permitted, calculations have been
performed in which the Macro Atom formalism has not been employed, and all
lines have been treated as resonance lines (no line branching). The comparison
of results using the full Macro Atom approach and the resonance 
scattering assumption forms the basis of the differential analyses 
presented in Sections 5 and 6.

Once the emission frequency of the packet is determined, the new direction
of propagation is determined by randomly sampling the
distribution function for the escape probability. Thus the probability of 
emission with direction cosine $\mu$ is 

\begin{equation}
P(\mu) \; \mbox{d}\mu  = K \frac{1 - e^{-\tau_{s}(\mu)}}{\tau_{s}(\mu)}\; \mbox{d}\mu
\end{equation}
where $K$ is a normalisation constant.

\subsection{The mass-loss rate}

Once all the packets have propagated through the model and have either 
escaped from the upper boundary or re-entered the stellar core, the
rate of energy deposition in the wind can be computed. It is given by
AL85

\begin{equation}
L_{T} = L_{*} \frac{ \sum_{k} \epsilon_{k} - \sum_{j} \epsilon_{j}
- \sum_{i} \epsilon_{i}}{\sum_{k} \epsilon_{k} - \sum_{j} \epsilon_{j}}
\end{equation}
where the $\epsilon$'s are the rest frame packet energies and the sums run
over all packets as they enter the calculation (index $k$), escape the
wind (index $i$) and re-enter the star (index $j$). 
This energy goes into lifting 
material out of the gravitational field of the star and so the mass-loss 
rate implied by the Monte Carlo calculation
is given by (LA93)

\begin{equation}
\Phi_{\mbox{\scriptsize MC}} = \frac{2 L_{T}}{v_{\infty}^2 - v_{c}^2 + v_{\mbox{\scriptsize esc}}^2}
\end{equation}
where $v_{\mbox{\scriptsize esc}}$ is the escape speed at $r=R_{c}$. This
can be compared to the mass-loss rate assumed in the modelling ($\Phi$). 
By performing
calculations for models with different values of $\Phi$,
$\Lambda = \Phi_{\mbox{\scriptsize MC}}/\Phi$
can be found as a function of $\Phi$ and so the self-consistent value of
$\Phi$ (i.e. when $\Lambda=1$) can be identified.

This approach 
does not look for local dynamical consistency between the driving force
and the velocity law: the mass-loss rate is determined algebraically
from the global requirement $\Lambda=1$ rather than as an eigenvalue at 
a critical point in the flow. LA93 used this method and
showed that, for their W-R model, the Monte Carlo radiative line acceleration
differed locally from that implied by the assumed velocity law by up 
to 0.29~dex. Ultimately models with realistic calculations of the line
acceleration which impose local dynamic consistency are desirable but, as
discussed by LA93, in view of the other simplifications made,
to pursue local consistency here is not warranted.

\section{Atomic Data}

Atomic models and radiative transition probabilities are needed for the 
computation of the mass-loss as described above. For the calculations presented
here, the atomic data has all been taken from the Kurucz \& Bell (1995) atomic
database. All elements with solar abundances $X/H < 3 \times 10^{-8}$
have been neglected. For the elements which are included all ionisation
stages up to {\sc vi} that are in the database are used (higher ionisation
species are not important for the OB-star models below).
The elements and ionisation stages from which lines
have been included in the
calculations presented in Section~5 are listed in Table~1.

\begin{table}
\caption{The elements and ions included in the line list.}
\begin{tabular}{lccc} \hline
Element& Ions& Element & Ions\\ \hline
H&{\sc i}& He& {\sc i,ii}\\
C& {\sc i -- iv}     & N &{\sc i -- vi}\\
O &{\sc i -- vi} & F&{\sc i -- vi} \\
Ne &{\sc i -- vi} & Na&{\sc i -- vi} \\
Mg &{\sc i -- vi} & Al& {\sc i -- vi}\\
Si &{\sc i -- vi} & P&{\sc i -- vi} \\
S &{\sc i -- vi} & Cl &{\sc i -- v}  \\
Ar & {\sc i -- v}& K &{\sc i -- v} \\
Ca &{\sc i -- vi} & Ti& {\sc i -- vi}\\
Cr &{\sc i -- vi} & Mn &{\sc i -- vi}\\
Fe &{\sc i -- vi} & Co &{\sc i -- vi}\\
Ni & {\sc i -- vi}& & \\ \hline
\end{tabular}
\end{table}

For each ion a cut-off in $\log gf$ was made to exclude weak transitions (in 
order that the line list is of a manageable size). Atomic models were then 
constructed for each ion, based on its line list and a complete, frequency
ordered line list assembled for the computations. In total, around
$2.5 \times 10^5$ transitions are included in the line list.

\section{OB stars}

In this section, calculations of mass-loss rates for O and B stars will be
presented. The models used here have all been constructed using a 
velocity law (equation 1) with $\beta=1$, $v_c =0$ and the density
determined from equation (2). 
In all cases the core radius ($R_{c}$) is determined from the luminosity ($L$)
and effective temperature ($T_{\mbox{\scriptsize eff}}$) using the
Stefan-Boltzmann equation.
The electron temperature, $T_e$ is assumed to
be constant throughout each model and is set by

\begin{equation}
T_{e} = 0.9 T_{\mbox{\scriptsize eff}} \; \; .
\end{equation}

For their model of $\zeta$~Puppis, AL85 adopted 
a constant radiation temperature somewhat lower than the effective
temperature of the star. A similar approach is used here. In principle
the radiation temperature for each ion may be determined iteratively
during the Monte Carlo simulation (see e.g. Lucy 1999a). 
Iterative calculations of $T_{R}$ were performed using
the $\zeta$~Pup model discussed below in which the radiation temperature
for each ion was computed using the simulated radiation field above
its ground state ionisation edge. 
The results of these calculations
could be adequately described by a constant value of
$T_{R}$ for 
ionisation edges below that of He~{\sc ii} and no radiation above this
edge owing to the truncation of the incident radiation field at 228~\AA.
Therefore, to avoid the need for iteration, in the OB star calculations 
that follow

\begin{equation}
T_{R} = T_{\mbox{\scriptsize eff}} - 9,000\mbox{~K}
\end{equation}
is adopted for ions with ionisation potentials less that He~{\sc ii}
and no ionisation to higher ions is included.
The decrement of $9,000$~K is larger than the decrement adopted
by AL85 and was chosen in order to give reasonable agreement
between the models presented here and the more sophisticated models used
by VdKL00.

\subsection{$\zeta$ Pup}

The first model that will be used to show the differential effect of 
line branching on mass-loss calculations is chosen to represent the well
studied O4~I~(f) star, $\zeta$~Puppis. The adopted stellar parameters are 
taken from Puls et al. (1996) and listed in Table~2. 
There is significant uncertainty in the mass: the Puls et al. value was 
derived 
spectroscopically rather than from evolutionary calculations 
(evolutionary masses are typically higher -- 
e.g. 70 M$_{\odot}$ as adopted by Lamers \& Nugis 2003).
The same element
composition used by AL85 is assumed: i.e. enhanced helium,
but all other abundances assumed to be solar.

\begin{table}
\caption{Stellar parameters used for $\zeta$~Puppis. From 
Puls et al. (1996).}
\begin{tabular}{lc}\hline
Parameter & Value\\\hline
Mass, $M_{*}$ & 52.5 M$_{\odot}$\\
 Luminosity, $L_{*}$& $10^6$ L$_{\odot}$\\
Effective temperature, $T_{\mbox{\scriptsize eff}}$ & 42,000 K\\
Terminal velocity,
$v_{\infty}$ & 2250 km~s$^{-1}$\\ \hline
\end{tabular}
\end{table}

To investigate the effect of line branching on the mass-loss rate of 
$\zeta$~Puppis a range of models with different input values of $\Phi$
have been used to compute $\Lambda(\Phi)$ (see Section 3.4), 
initially including
the complete treatment of line branching. Separate calculations were then
performed to find $\Lambda(\Phi)$ when line branching is neglected and
all line scatterings are treated as resonant scattering. Fig.~1 shows
$\Lambda(\Phi)$ versus $\Phi$ for both sets of calculations.

\begin{figure}
\epsfig{file=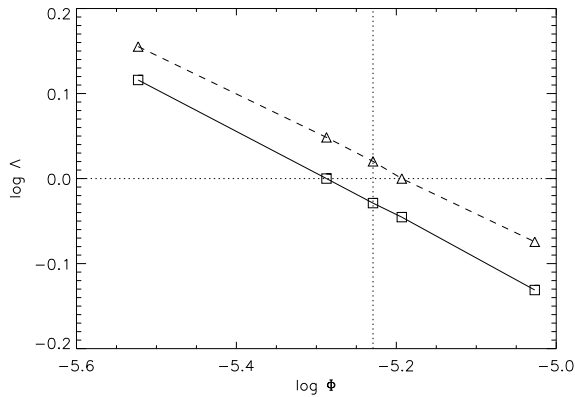,width=8cm}
\caption{$\Lambda$ as a function of $\Phi$ for $\zeta$~Puppis: 
the boxes (connected by solid
line) show computations with line branching and the triangles (connected 
with dashed line) show computations with resonance line scattering only.
The dotted horizontal line is $\Lambda=1.0$ (self-consistent model) and the
vertical dotted line indicated the observed mass-loss rate (from
Puls et al. 1996). The $\Lambda$ values are accurate to around
$\pm 0.01$~dex.}
\end{figure}

The calculations yield a self-consistent ($\Lambda=1$) mass-loss rate for 
$\zeta$~Puppis
of $5.16 \times 10^{-6}$~M$_{\odot}$~yr$^{-1}$ if line branching is included
and $6.41 \times 10^{-6}$~M$_{\odot}$~yr$^{-1}$ if only resonance scattering
is permitted. A sufficient number of packets were used that these values of 
$\Phi$ are correct to better than $\pm 2$ per cent.
Thus the differential effect of line branching in the $\zeta$~Puppis model
is a reduction in the mass-loss rate by around 20 per cent.

This reduction occurs as a result of photon leakage: 
the net effect of line branching 
is the transfer of packets from frequency ranges with many strong lines
to regions with fewer lines.
This
reduces the average number of lines with which a packet interacts and thus
reduces the mass-loss rate. This effect is illustrated by Fig.~2 which 
shows the number of line scatterings that packets undergo when line branching
is allowed compared to when it is not. Clearly, line branching significantly
reduces the number of packets that undergo multiple ($> 10$) line
scattering events. Of course the wind is sufficiently diffuse that, in both 
cases, the radiative acceleration is dominated by the large number 
of packets which undergo relatively few scattering and so the differential
effect is relatively small (the wind performance number for the $\zeta$~Puppis
model is only $\eta = 0.56$). Clearly, however, the effect will be
bigger when multiply scattered packets become more dominant: i.e. when 
$\eta > 1$. 

\begin{figure}
\epsfig{file=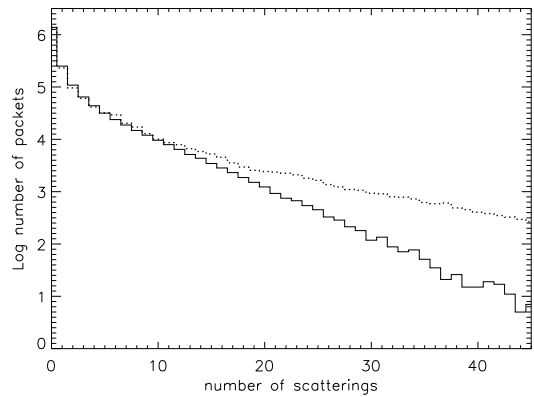,width=8cm}
\caption{The number of packets versus the number of line scatterings for 
$\zeta$~Puppis from Monte Carlo simulations with a total of $2 \times 10^6$
packets. The solid histogram shows a simulation where line branching is 
allowed.
The dotted histogram is the case where only resonance scattering is 
included.}
\end{figure}

Fig.~3 shows the fractional contribution to the energy deposition rate from 
all the ions included in the $\zeta$~Puppis modelling.
These were computed by summing the 
contributions to the energy deposition rate from 
line scattering for each ion
during the Monte Carlo simulation, not by performing separate
computations with lines of only one ion at a time.
The results shown are from a calculation in which branching was included.
For $\zeta$~Puppis the global mass-loss rate has significant contributions
from Fe and from several of the light elements (S, Ar, N, O, C).
When branching is not permitted the relative contributions of the elements
are very similar except that iron contributes very slightly more:
this is expected since the iron group
elements have the most complex spectra with many bunches of lines at
similar wavelengths meaning that they are the most sensitive to photon
leakage.

\begin{figure}
\epsfig{file=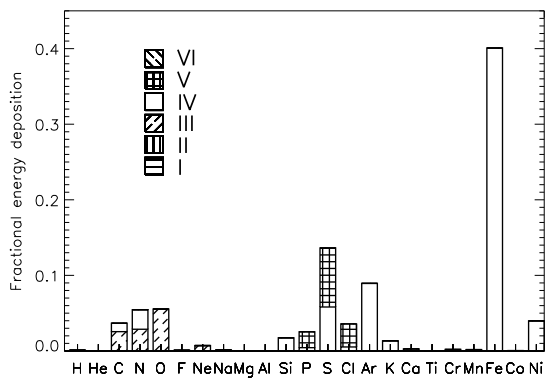,width=8cm}
\caption{The relative importance of the different elements/ions to the
mass-loss-rate for $\zeta$~Puppis. 
The chart shows the fraction of energy deposited in the wind by each element
in a Monte Carlo simulation including branching. The fractions are
not significantly different if branching is neglected.
The bars for each element are subdivided by ionisation stage.
Note that the fractions do not sum to 1.0 because of the contribution of
scattering by free electrons (not shown).
}
\end{figure}

These results must be put into context regarding their 
relevance to our understanding
of massive stars. Firstly, let us compare the derived mass-loss rates for
$\zeta$~Puppis with those derived from models made by others. AL85 
adopted a slightly smaller stellar mass (40~M$_{\odot}$) 
and higher terminal velocity (2600~km~s$^{-1}$) than used 
here and obtained a mass-loss rate for $\zeta$~Puppis of 
$5.2 \times 10^{-6}$~M$_{\odot}$~yr$^{-1}$ from their standard model,
with values ranging from about 4 -- 7~$\times 10^{-6}$~M$_{\odot}$~yr$^{-1}$
from their models with different treatments of ionisation and assumed
velocity law. This is in good agreement with the mass-loss rate
derived here with resonant scattering only (AL85 did not
attempt to include line branching in their calculations).
Using the parameters from Table~2, the VdKL00 mass-loss recipe
gives a mass-loss rate of $9.4 \times 10^{-6}$~M$_{\odot}$~yr$^{-1}$, 
similar to the value quoted for $\zeta$~Puppis by VdKL00.
This is a little higher than the mass-loss rate obtained here, 
but it is important to note (see Fig.~1) that $\Lambda$ is a
weak function of $\Phi$ and thus relatively small differences in the modelling
approach (computation of $\Lambda$) lead to larger differences in the
derived globally self-consistent value of $\Phi$.
If $\Phi=9.4 \times 10^{-6}$~M$_{\odot}$~yr$^{-1}$ is used in the
modelling here, a Monte Carlo simulation 
(with resonance scattering only) leads to
$\Lambda =0.8$, indicating that the models used here reproduce 
the VdKL00 energy deposition 
rates adequately, to within 20 per cent.
Given that many simplifying assumptions have been made here (e.g. 
the stellar radiation field is assumed to be black-body, 
a single value of $T_{r}$ has been adopted for all ions at all radii,
no bound-free 
or free-free processes are included, 
a core-halo approach is adopted in the modelling), 
some discrepancy with the
VdKL00 results is expected.

\begin{table*}
\caption{Computed mass-loss rates for a selection of OB-star models. 
$\Phi_{\mbox{\scriptsize res}}$ and $\Phi_{\mbox{\scriptsize bran}}$
are the mass-loss rates computed with resonance scattering only
and with line branching, respectively.
$\Phi_{\mbox{\scriptsize VdKL}}$ is the mass-loss rate from the 
VdKL00 recipe. The terminal velocity ($v_{\infty}$) is taken as
1.3 times the effective escape speed 
($v^{\mbox{\scriptsize eff}}_{\mbox{\scriptsize esc}}$) on the cool
side of the bi-stability jump 
and 2.6 times 
$v^{\mbox{\scriptsize eff}}_{\mbox{\scriptsize esc}}$
on the hot side (VdKL00). $\eta$ is the wind performance number,
$\eta \equiv \Phi v_{\infty} c / L_{*}$. 
$\bar{\rho}= \Phi_{\mbox{\scriptsize bran}}/(4\pi R_{c}^2 v_{\infty})$ is the
characteristic mass density in the model.
The results are accurate to $\sim 
\pm 2$ per cent.}
\begin{tabular}{lccccccccccc}\hline
Model&$\log L_{*}$ & $M_{*}$ & $T_{\mbox{\scriptsize eff}}$ & $v_{\infty}$($v_{\infty}/v^{\mbox{\scriptsize eff}}_{\mbox{\scriptsize esc}}$)& $\log \Phi_{\mbox{\scriptsize bran}}$ &
$\eta_{\mbox{\scriptsize bran}}$&
$\log \Phi_{\mbox{\scriptsize res}}$ &
$\eta_{\mbox{\scriptsize res}}$&
$\log \Phi_{\mbox{\scriptsize VdKL}}$ &
$\Phi_{\mbox{\scriptsize bran}}/\Phi_{\mbox{\scriptsize res}}$&
$\log \bar{\rho}$\\
&($L_{\odot}$)&($M_{\odot}$)&(K)& (km~s$^{-1}$) & ($M_{\odot}$ yr$^{-1}$)&
& ($M_{\odot}$ yr$^{-1}$)&& ($M_{\odot}$ yr$^{-1}$)&
&(g cm$^{-3}$)
\\ \hline
A&6.0&60&50,000& 2606 (2.6)& -5.01 &1.24 & -4.89& 1.64 & -5.05 & 0.75 &-12.67\\
B&6.0&80&50,000& 3265 (2.6)&-5.23 & 0.94 & -5.14 &1.12 &  -5.21 & 0.82&-12.99\\
C&6.0&60&40.000& 2085 (2.6)& -5.29 & 0.52 & -5.17 & 0.68 & -5.04 &0.76&-13.24\\
D&6.0&80&40,000& 2612 (2.6)& -5.51 & 0.39 & -5.42 & 0.49& -5.20& 0.81&-13.56\\
E&5.5&40&40,000& 2660 (2.6)& -6.24 & 0.23 & -6.19&0.26&-5.90 &0.89& -13.79\\
F&5.0&20&40,000& 2619 (2.6)& -6.95 & 0.14 & -6.93& 0.15 &-6.61& 0.94& -14.00 \\
G&6.0&60&30,000&1564 (2.6)&-5.75&0.13&-5.67&0.16&-5.32& 0.83& -14.07\\
H&5.0&20&30,000&1964 (2.6)&-7.03&0.09&-7.02&0.09&-6.89& 0.97&-14.45\\
I&6.0&60&20,000& 521 (1.3)& -4.54 & 0.72 & -4.37 & 1.07 & -4.58 &0.68&-13.09\\
J&6.0&80&20,000& 653 (1.3)& -5.19 & 0.20 & -4.67 & 0.68 & -4.75 & 0.30&-13.84\\
K&5.5&40&20,000& 665 (1.3)& -6.27 & 0.06 & -5.90& 0.13 &-5.45 &0.43& -14.43 \\ 
L&5.0&20&20,000& 655 (1.3)& -6.98 & 0.03 & -6.77 & 0.05&-6.15&  0.62& -14.62\\
M&6.0&60&20,000&1042 (2.6)&-5.78&0.08&-5.44&0.18&-5.06& 0.46& -14.63\\
N&5.0&20&20,000&1309 (2.6)&-7.30&0.03&-7.23&0.04&-6.63& 0.85&-15.25\\
O&6.0&60&15,000&391 (1.3)&-4.57&0.51&-4.47&0.64&-4.30$^a$& 0.80& -13.49\\
\hline
\end{tabular}

\noindent $^a$ This value was read from Fig 2. of VdKL00 since the
mass-loss recipe (section 5 of VdKL00) does not fit this model well.
\end{table*}

From observations of H$\alpha$, Puls et al. (1996) have derived a mass-loss 
rate of $5.9 \times 10^{-6}$~M$_{\odot}$~yr$^{-1}$ for $\zeta$~Puppis.
This value is indicated by the vertical line in Fig.~1, and is in good
agreement with the computed mass-loss rates discussed above. However, there is
significant uncertainty in the observational mass-loss rate: estimates from 
radio fluxes give only $2.4 \times 10^{-6}$~M$_{\odot}$~yr$^{-1}$ 
(Lamers \& Leitherer 1993). 
Clumping in the wind may affect the accuracy of these determinations
although this effect should be small for OB stars
(Kudritzki \& Puls 2000).
In view of the large observational uncertainty
in the mass-loss rate, the effect of line branching can be regarded as small
and, for the present, 
is not important for
modelling mass-loss from $\zeta$~Puppis to observable accuracy. 
However, the effect is not utterly negligible and 
upon a reduction in the observational uncertainty by a factor of a few, it
should become detectable (note, however, that since stellar winds show
intrinsic variability at the level of several tens per cent, time variability
would also need to be taken into account in reliable modelling of 
mass-loss rates to such high accuracy). 

\subsection{Grid of models}
\begin{figure*}
\epsfig{file=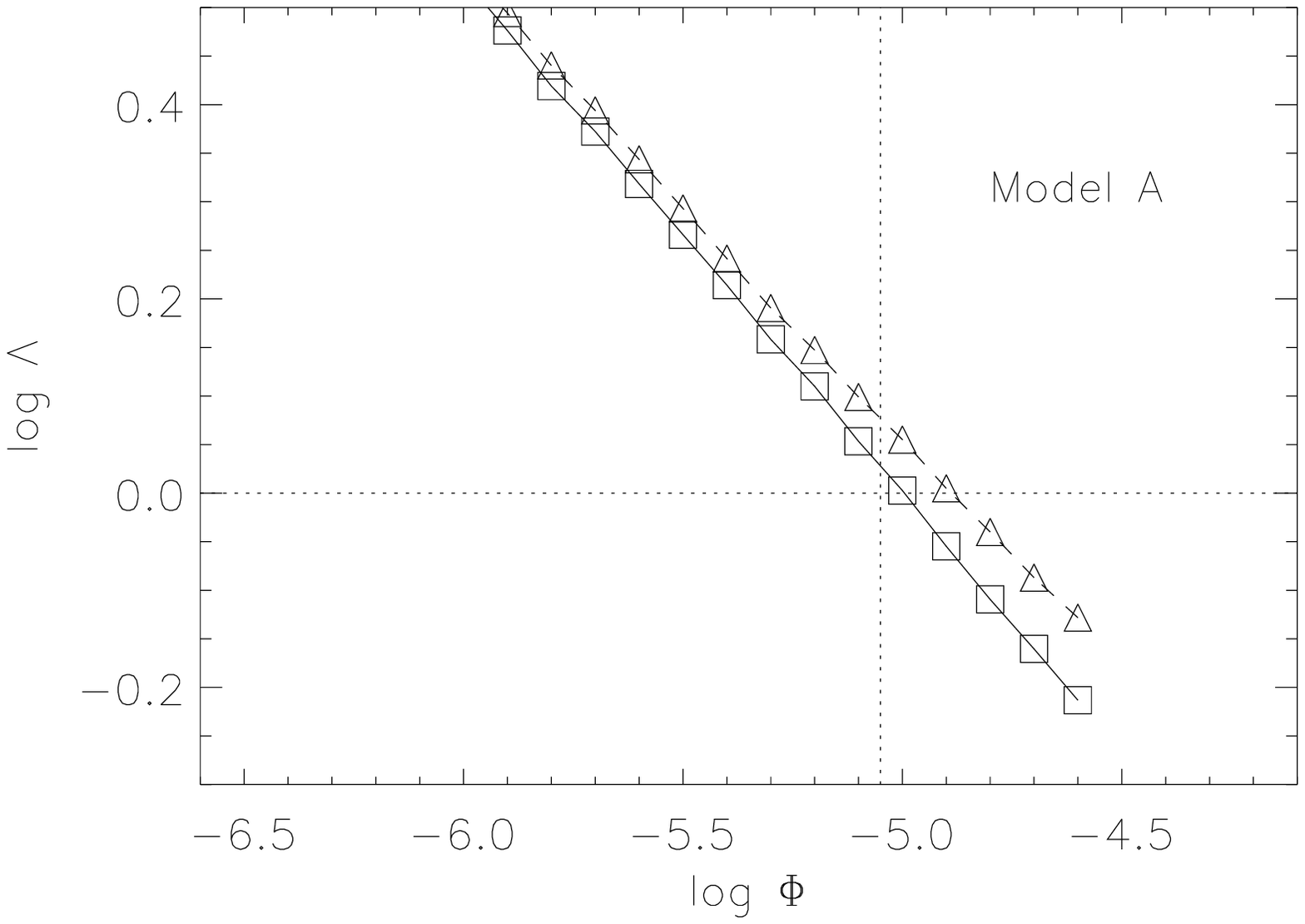,width=5.8cm}
\epsfig{file=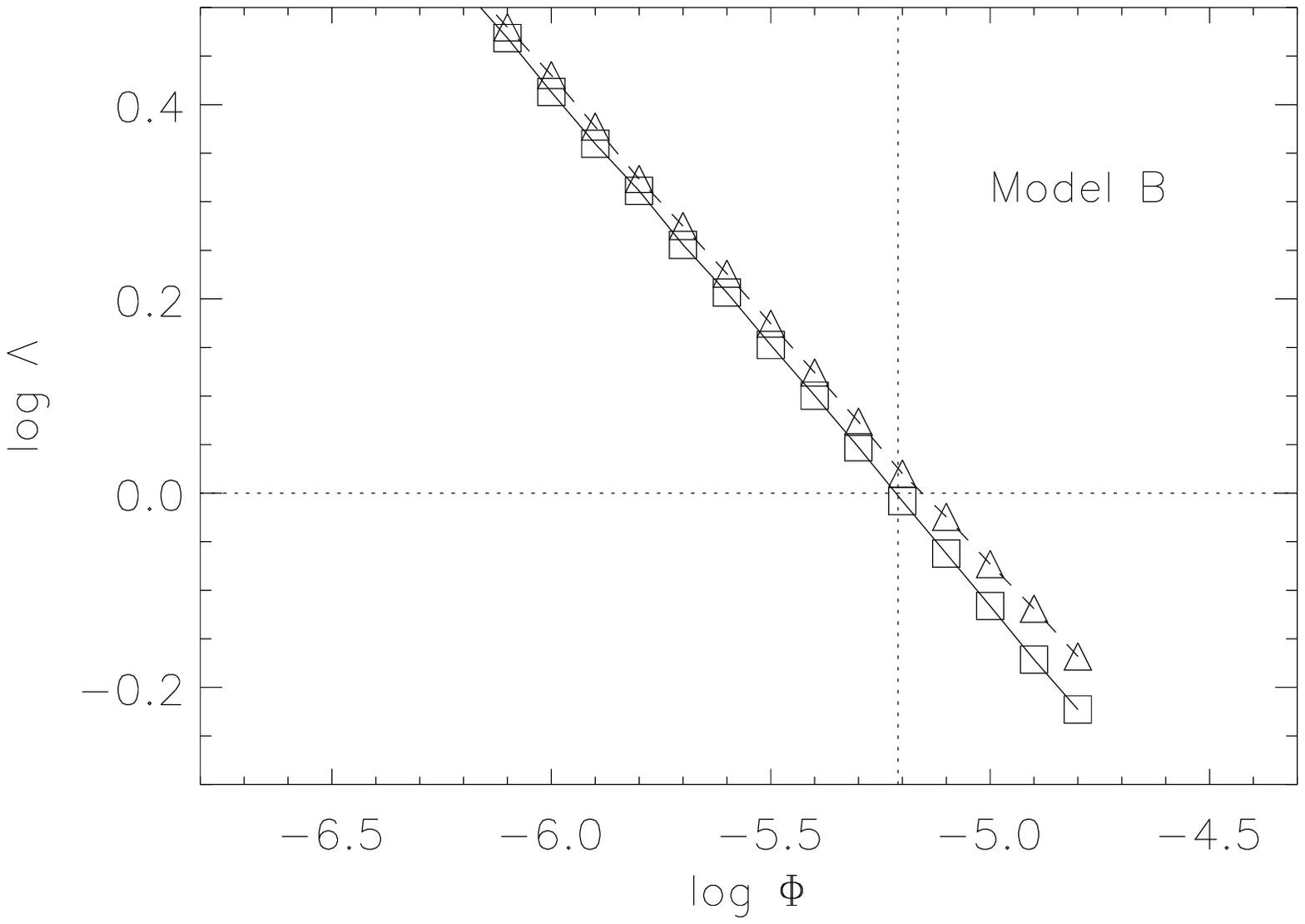,width=5.8cm}
\epsfig{file=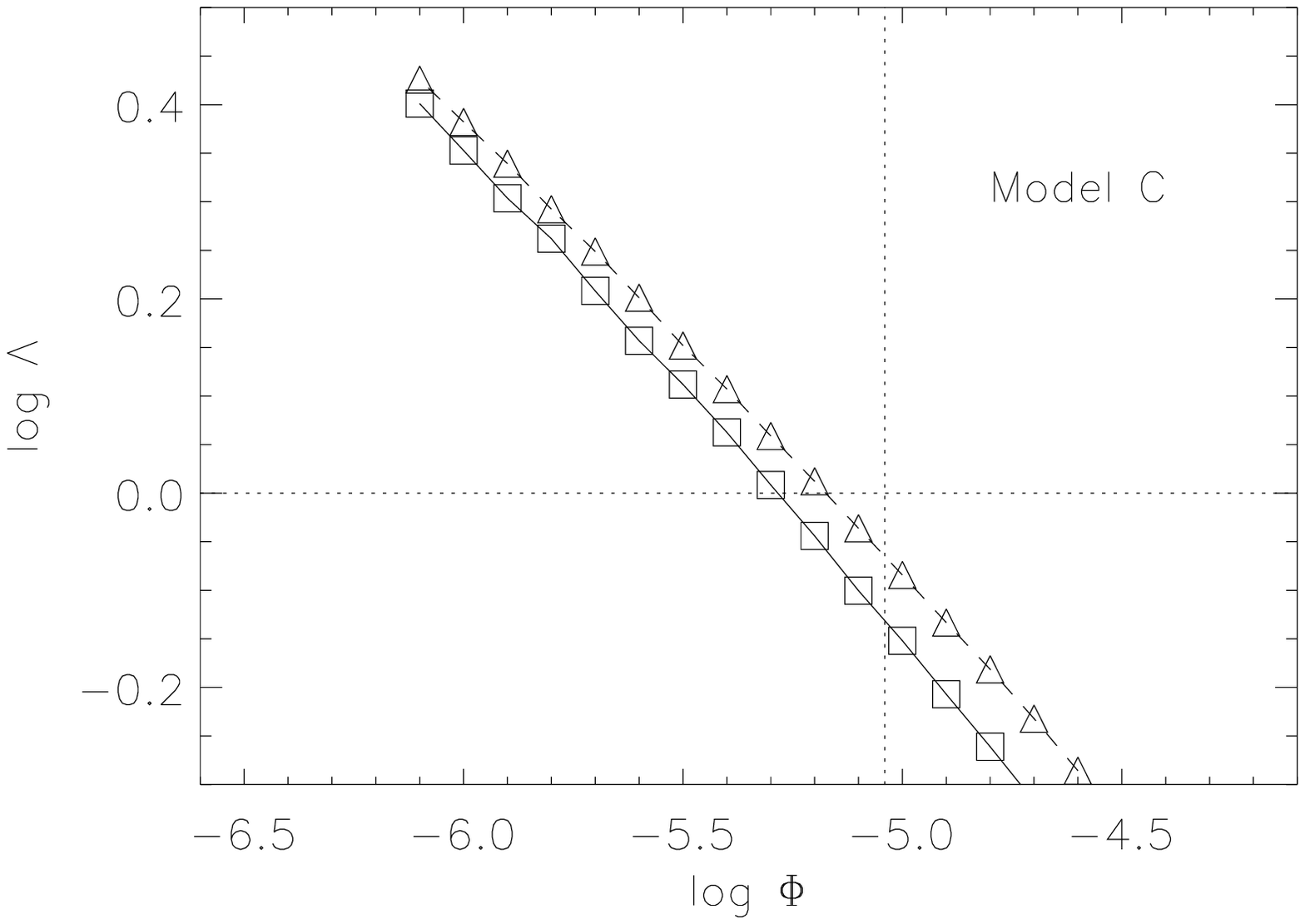,width=5.8cm}
\epsfig{file=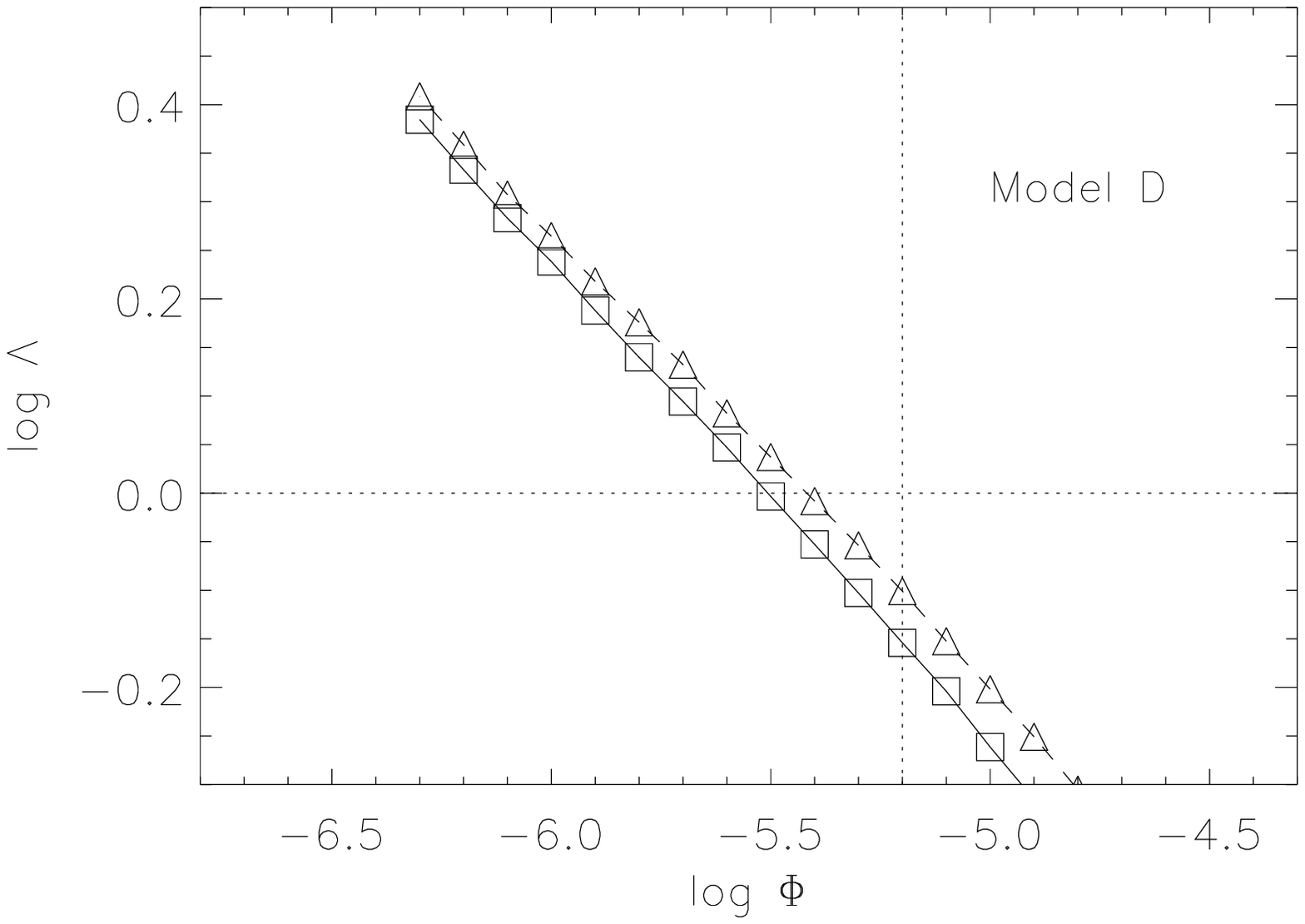,width=5.8cm}
\epsfig{file=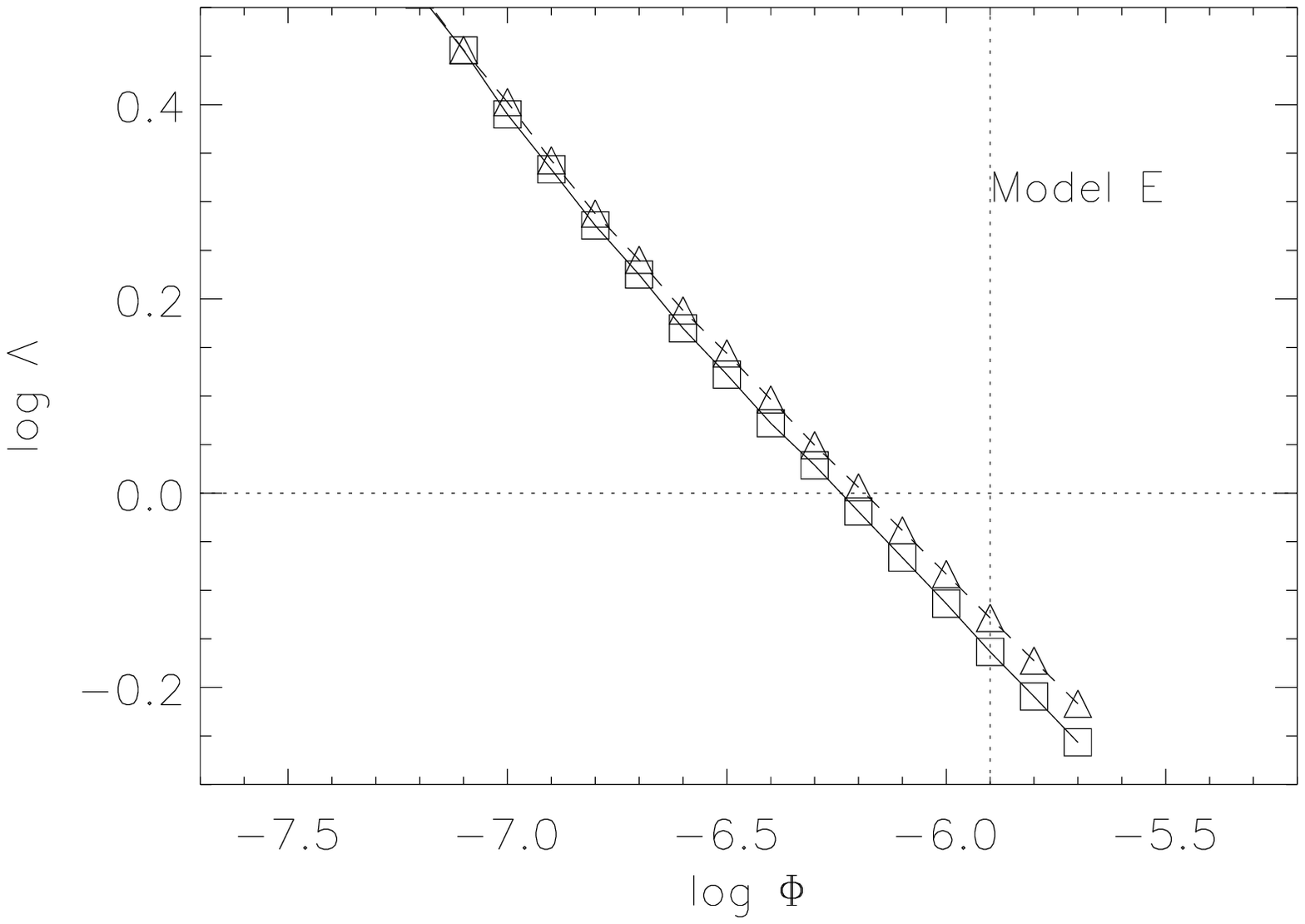,width=5.8cm}
\epsfig{file=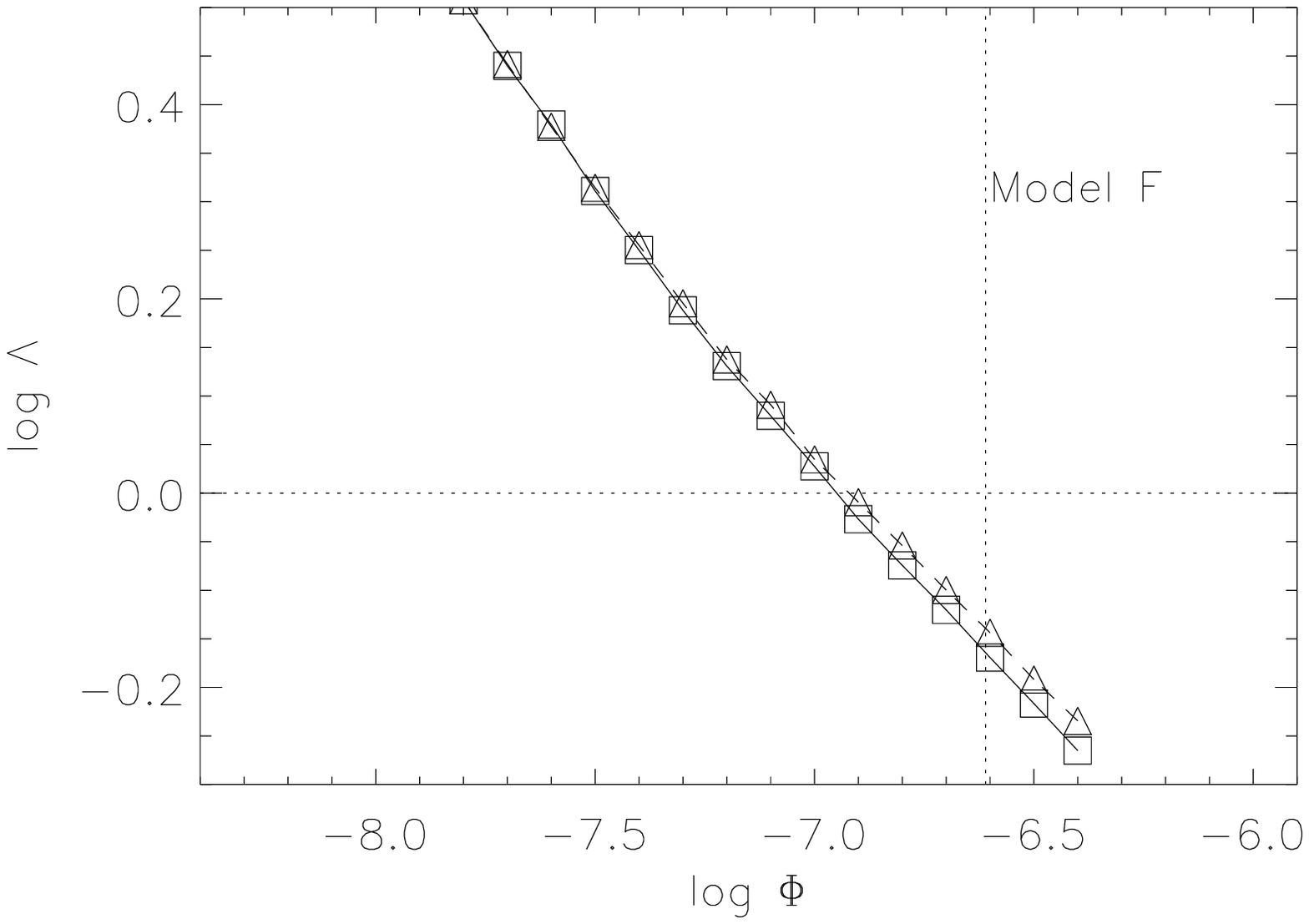,width=5.8cm}
\epsfig{file=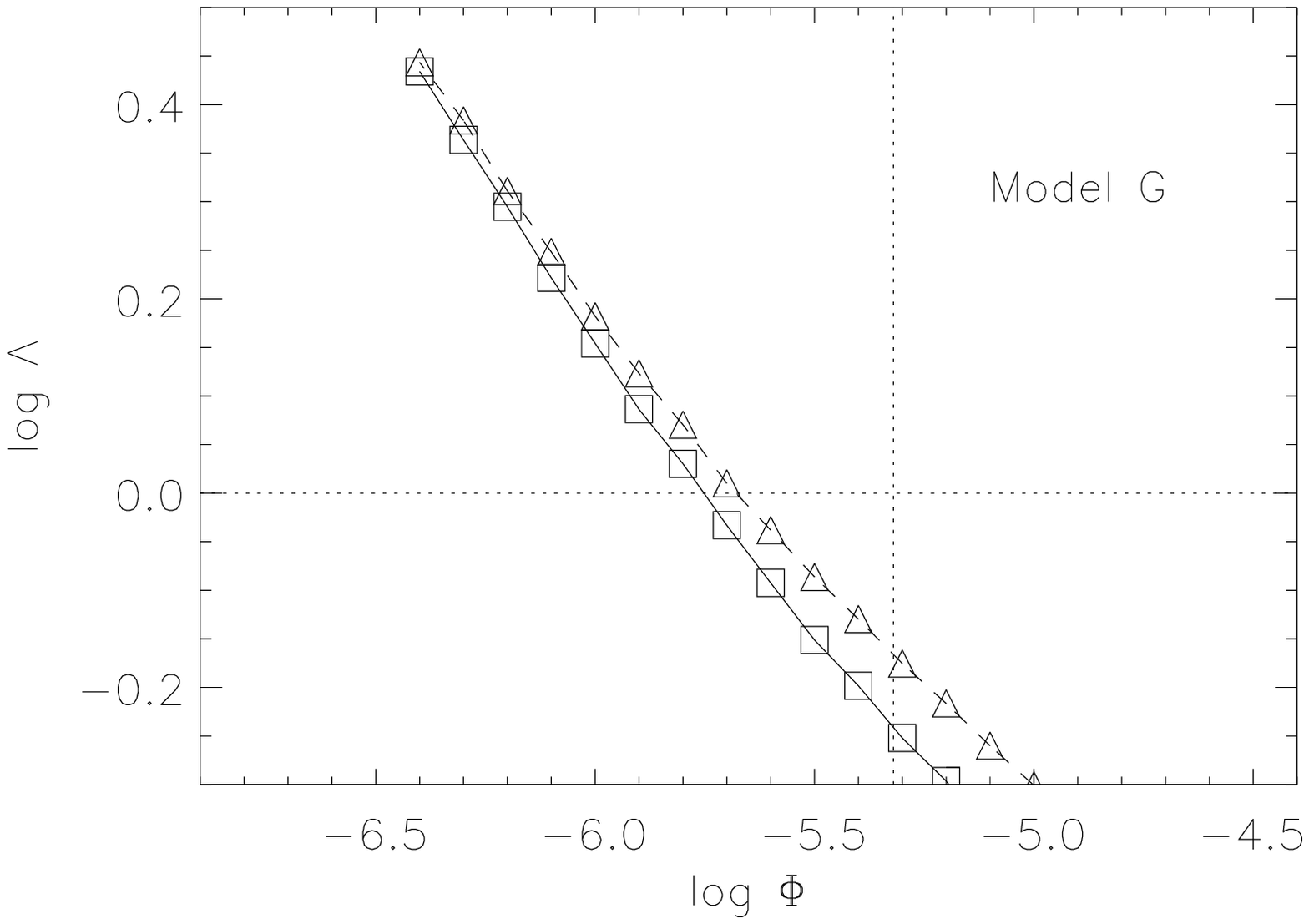,width=5.8cm}
\epsfig{file=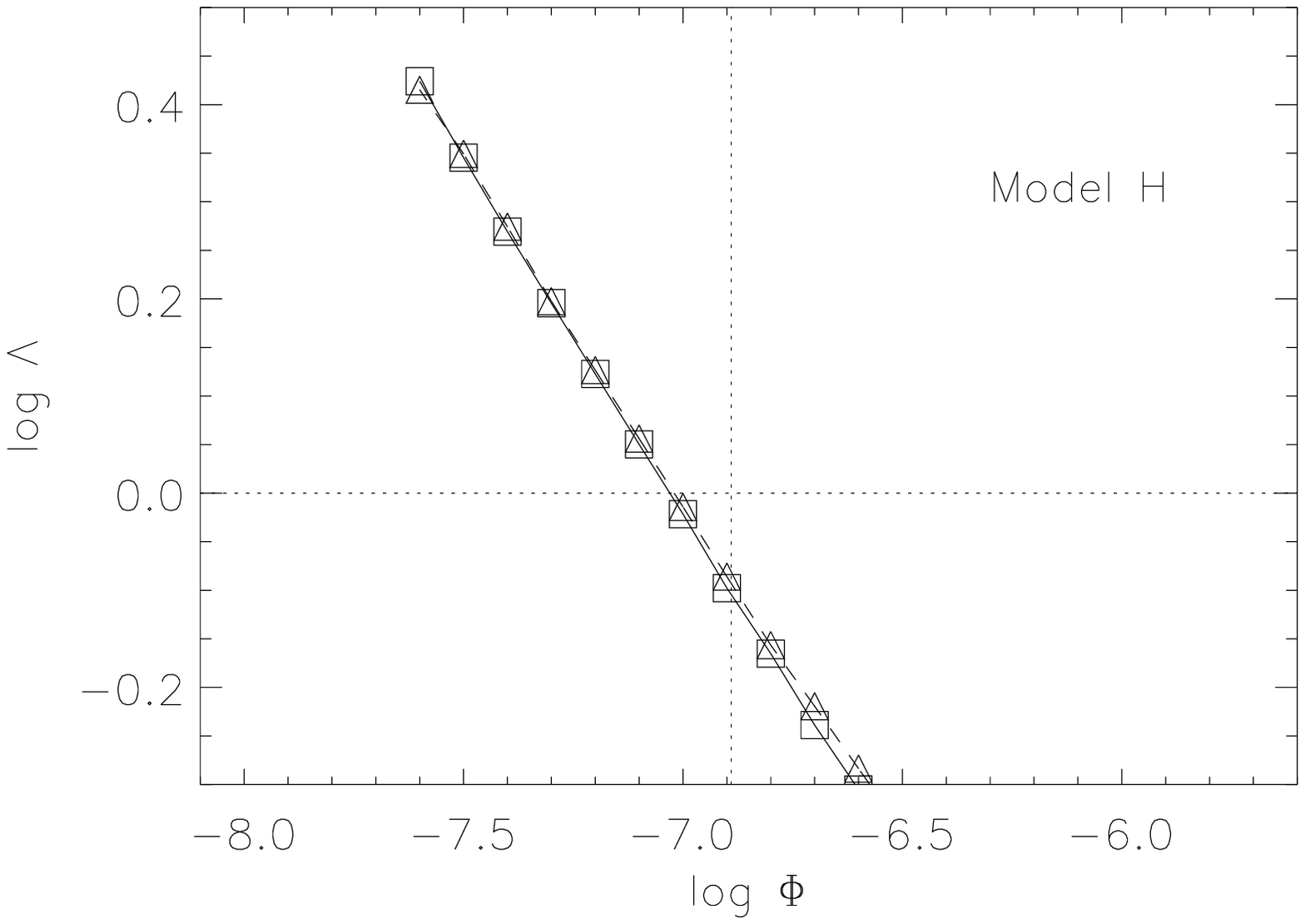,width=5.8cm}
\epsfig{file=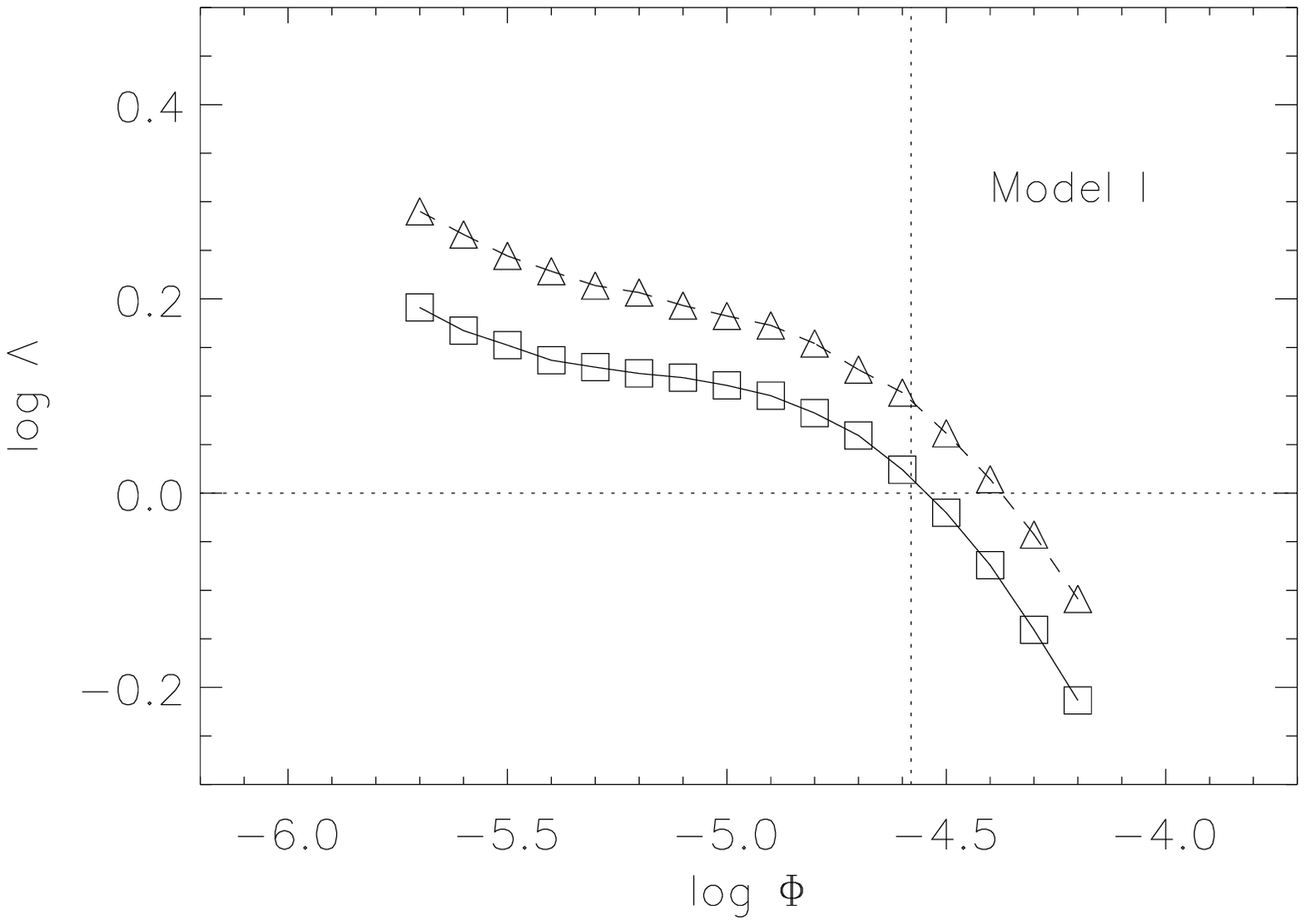,width=5.8cm}
\epsfig{file=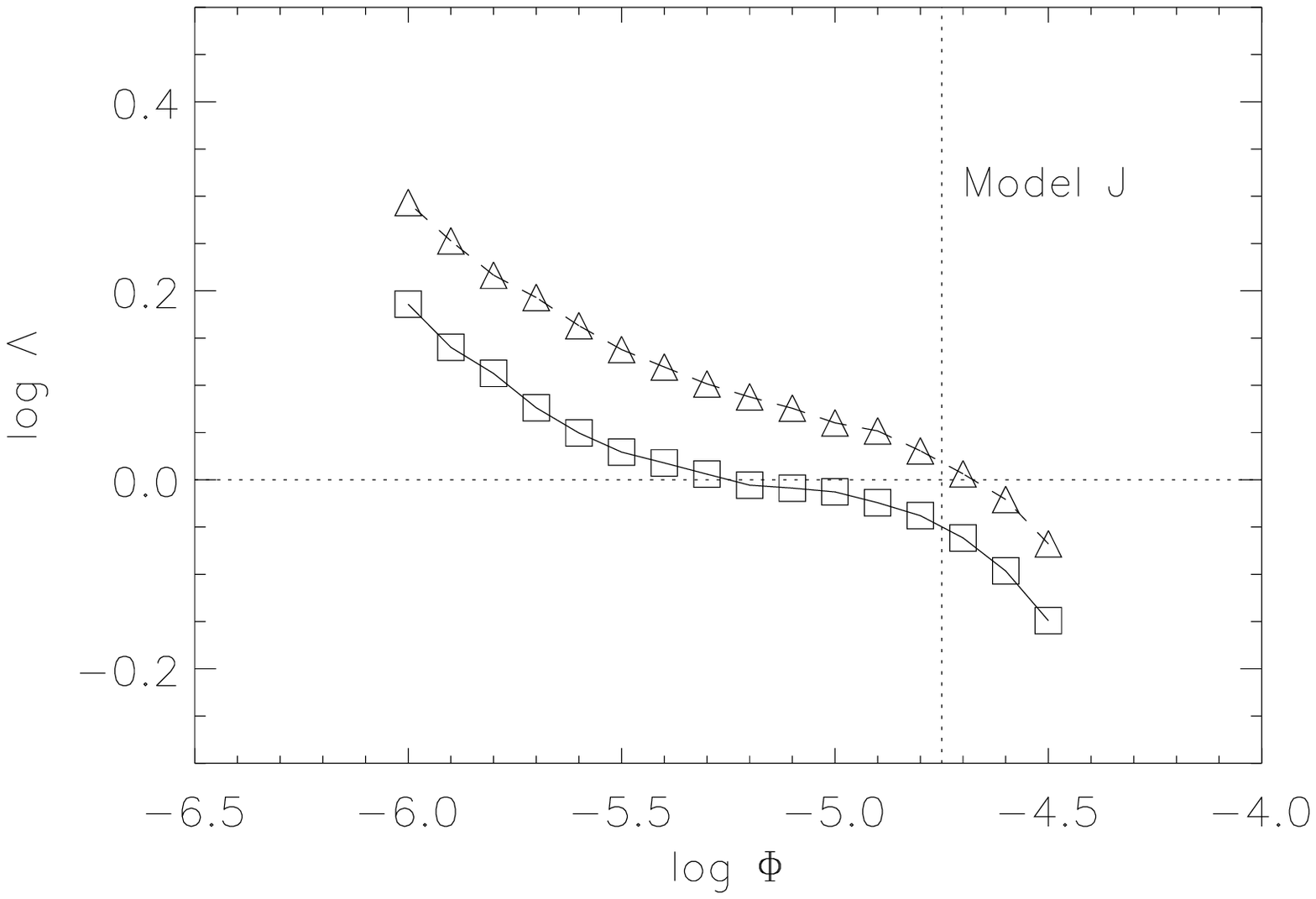,width=5.8cm}
\epsfig{file=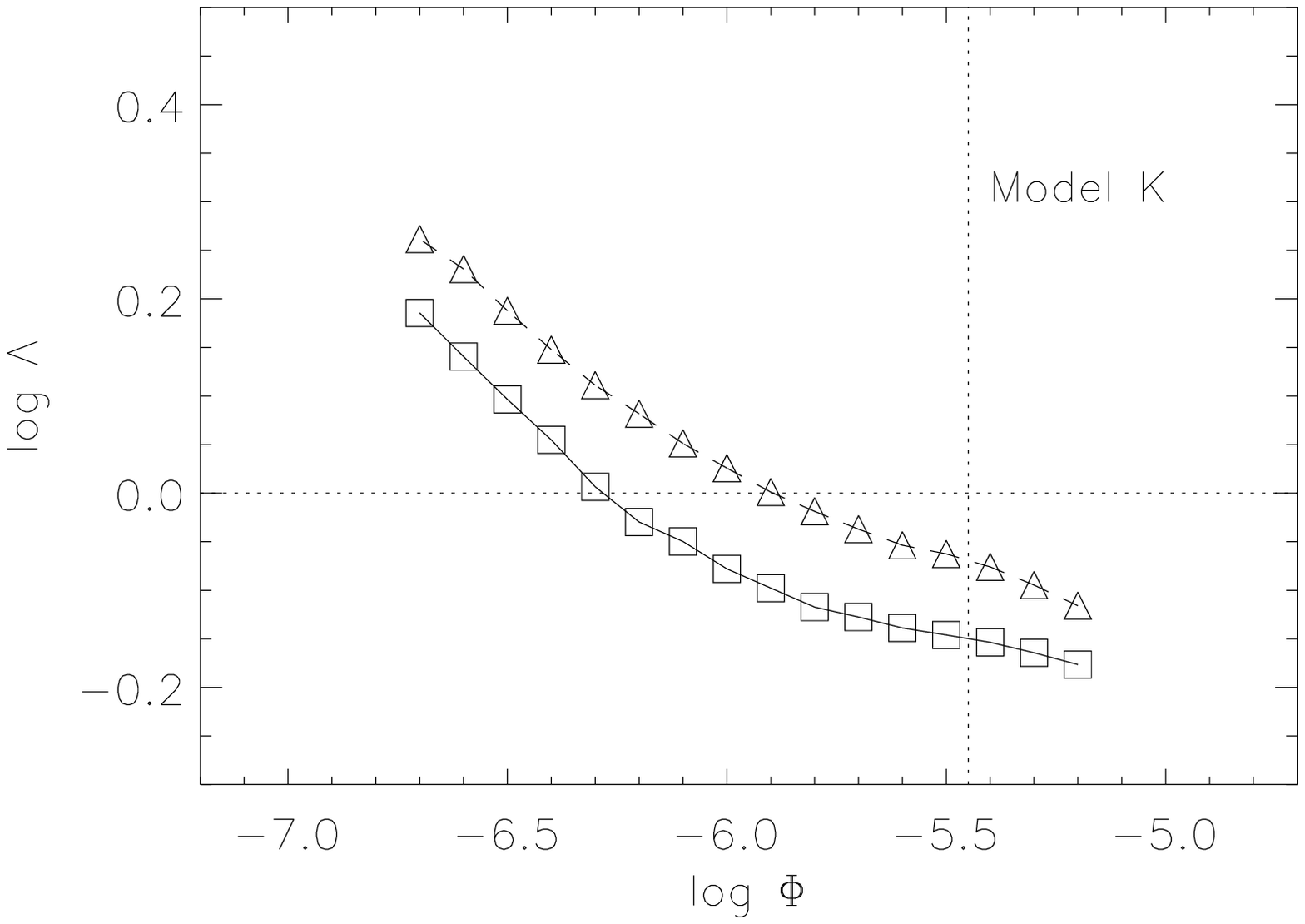,width=5.8cm}
\epsfig{file=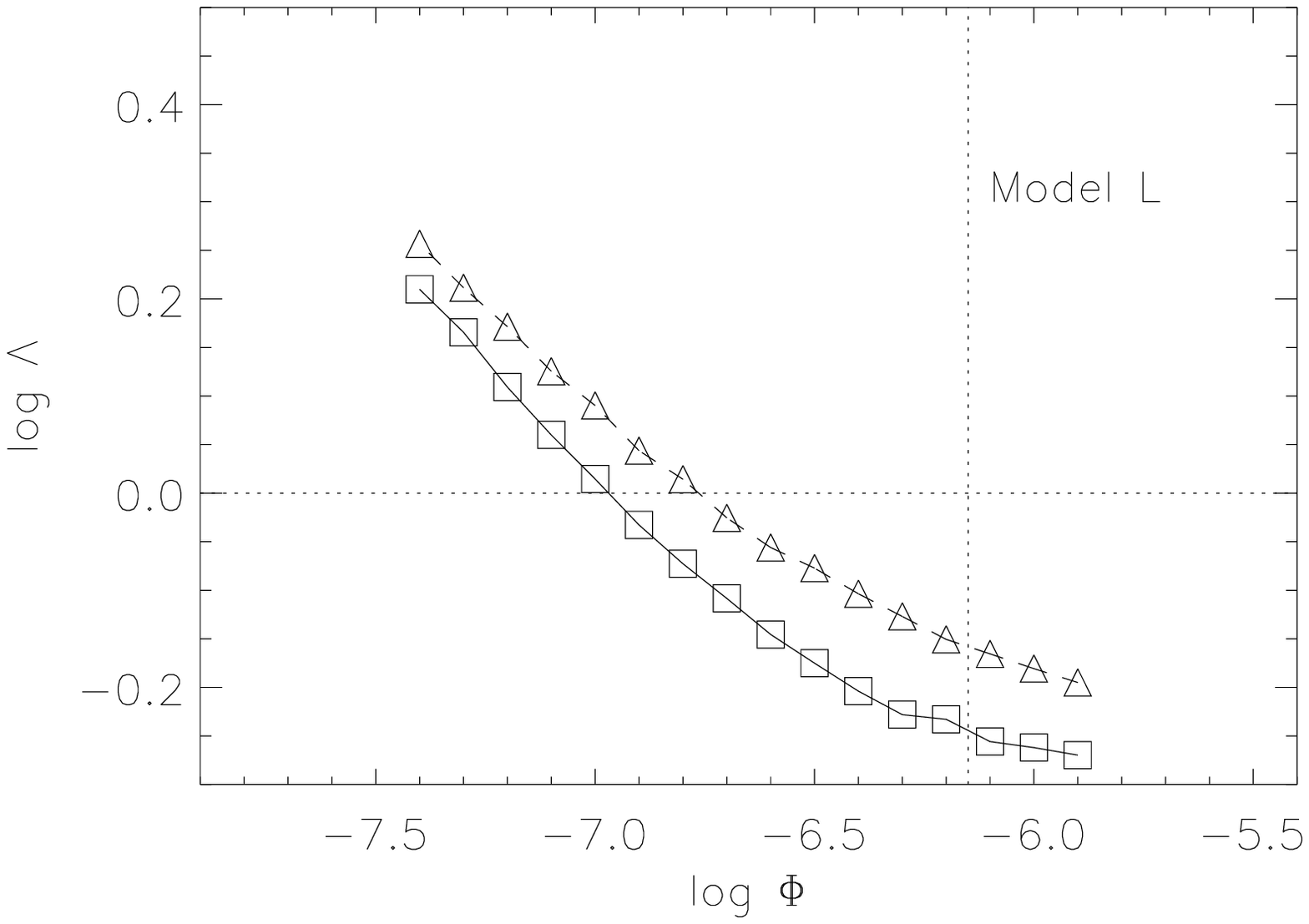,width=5.8cm}
\epsfig{file=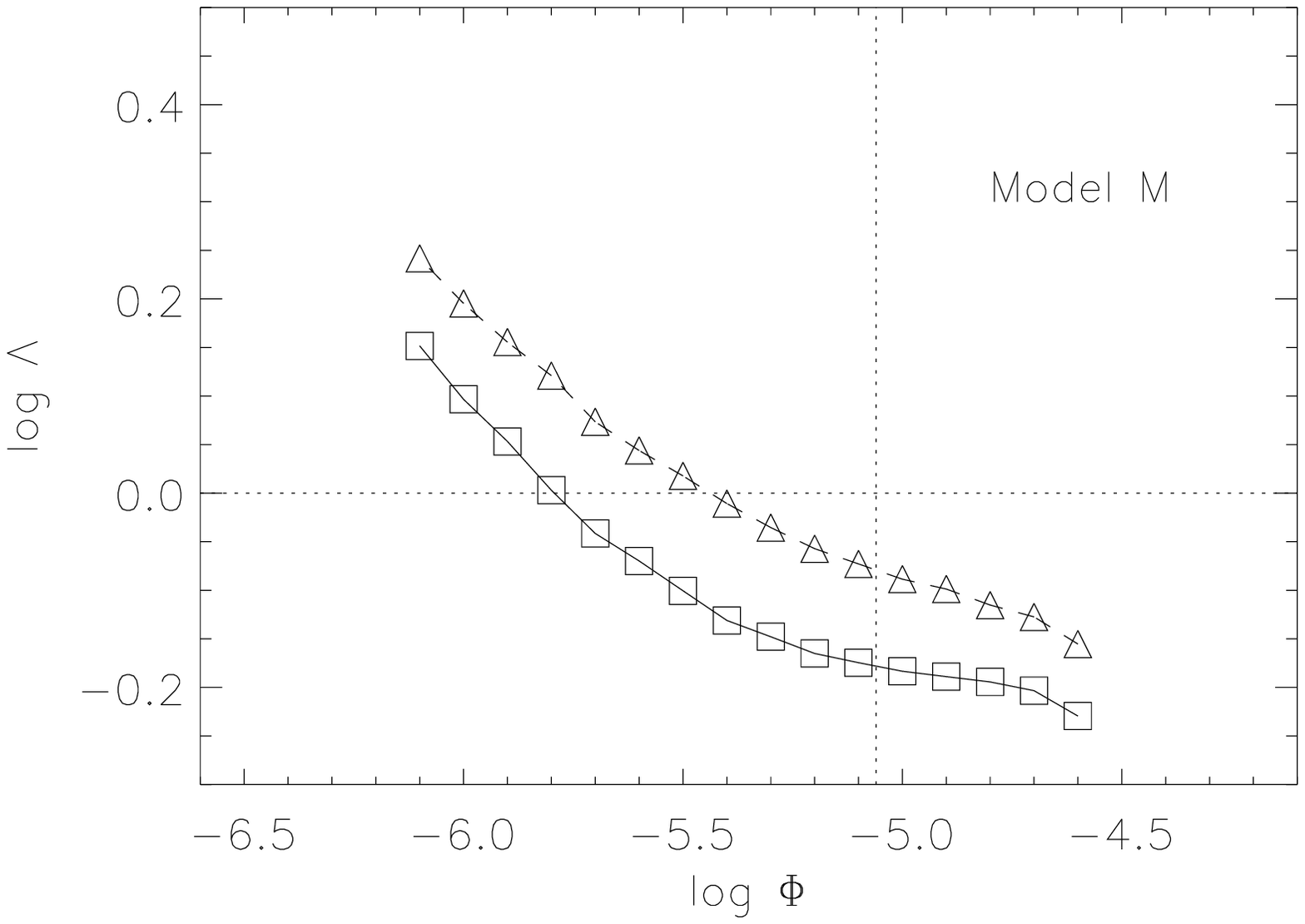,width=5.8cm}
\epsfig{file=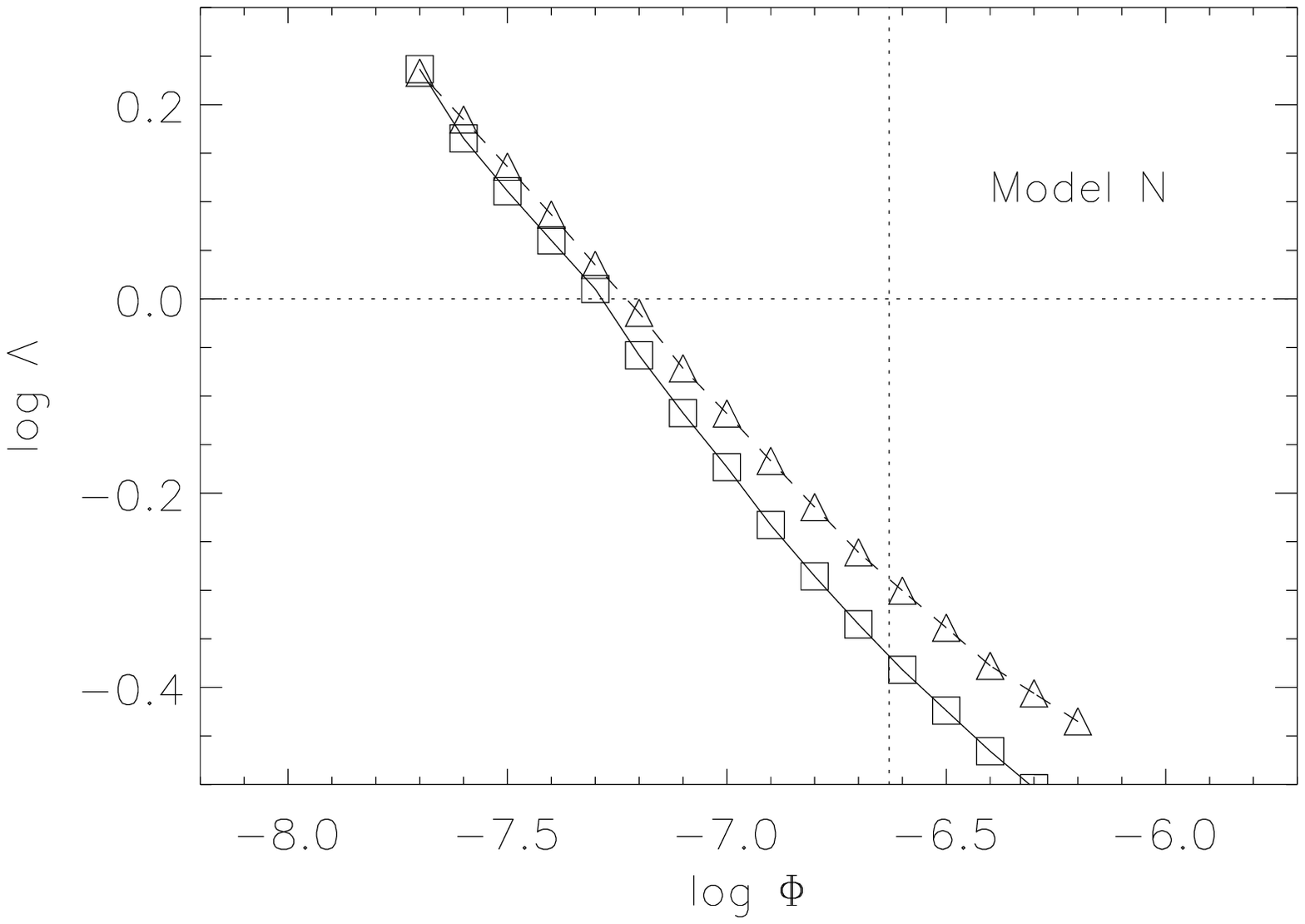,width=5.8cm}
\epsfig{file=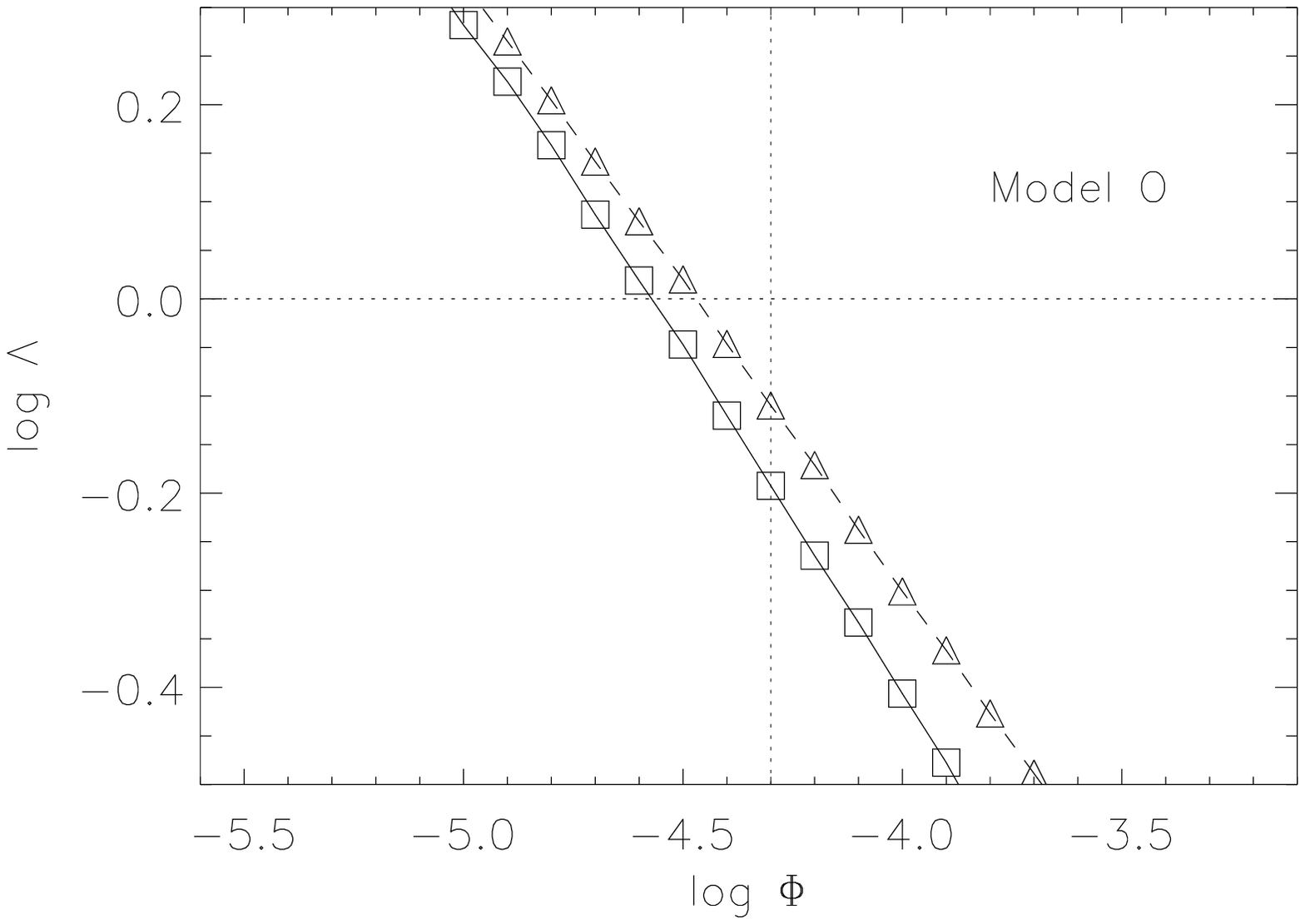,width=5.8cm}
\caption{$\Lambda(\Phi)$ for models A to O. In each case the boxes
(connected with solid lines) show the results with branching 
and the triangles (dashed lines) show the results without
branching. The horizontal dotted line shows $\Lambda=1.0$ 
(self-consistent model) and the vertical dotted line indicated the
mass-loss rate from the  VdKL00 recipe.}
\end{figure*}

Following on from the detailed discussion of $\zeta$~Puppis above, in this
section
the effects of line branching on the computed mass-loss rates are
investigated for a range of OB-star models spanning roughly the same parameter
space as the VdKL00 calculations. For each model
(defined by the stellar parameters $L_{*}$,$M_{*}$,$T_{\mbox{\scriptsize eff}}$
and
$v_{\infty}$), 
two $\Lambda(\Phi)$ curves have been computed, one with line branching
included and one without. The models considered are listed in 
Table~3 and the $\Lambda(\Phi)$ curves are shown for each model in 
Fig.~4.
For each model,
the globally self-consistent ($\Lambda(\Phi)=1$)
mass-loss rate 
was found both with and without line branching included 
($\Phi_{\mbox{\scriptsize bran}}$ and $\Phi_{\mbox{\scriptsize res}}$,
respectively).
These 
are given in Table~3. The Monte Carlo errors 
in the mass-loss rates are 
less than
2 per cent.
The table also gives the wind performance number
($\eta$) calculated with and without line branching, the VdKL00 
mass-loss rate ($\Phi_{\mbox{\scriptsize VdKL}}$)
for comparison, the ratio of the branching to 
resonance-only mass-loss rates and the 
characteristic density of the model 
$\bar{\rho}=\Phi_{\mbox{\scriptsize bran}}/(4 \pi R_{c}^2 v_{\infty})$.
The models are labelled A -- O, being ordered first by decreasing effective
temperature and secondly by decreasing $\bar{\rho}$.

Consider first the high temperature models appropriate for O stars
(A -- H). For these models the $\Lambda(\Phi)$ curves (Fig. 4) are all
approximately straight in $\log$-$\log$ space around $\Lambda=1$. The effect
of line branching (i.e. the vertical offset between the 
two curves in Fig. 4) is relatively small for all these models -- it leads
to a reduction in the self-consistent mass-loss rate 
by between three and 25 per cent, relative to calculations which neglect
branching.
At a given temperature (e.g. compare models C -- F) the reduction 
is closely correlated with the characteristic density 
($\bar{\rho}$). This is expected since the effect of line branching is related
to multiple scattering which is more common at
high densities.

Models I -- N all have $T_{\mbox{\scriptsize eff}} = 20,000$~K (early B 
stars). The first four of these models (I -- L) have terminal velocities 
which are typical 
for stars
cooler than the bi-stability jump ($v_{\infty} = 1.3 
v_{\mbox{\scriptsize esc}}^{\mbox{\scriptsize eff}}$ where
$v_{\mbox{\scriptsize esc}}^{\mbox{\scriptsize eff}}$ is the
effective escape velocity: see
VdKL99,00) while the other two have higher terminal velocities,
characteristic of stars hotter than the jump ($v_{\infty} = 2.6
v_{\mbox{\scriptsize esc}}^{\mbox{\scriptsize eff}}$).
For these models, the $\Lambda(\Phi)$ curves around $\Lambda=1$ 
are more complex than
for the higher temperature 
models discussed above (see Fig. 4): all the low temperature models
have $\Lambda=1$ close to a region in which $\Lambda(\Phi)$ passes
through a plateau. These plateaus are the result of changes in the driving
mechanism as a function of the density (which is
determined by the value of $\Phi$ adopted in the model)
and are thus physically related to the bi-stability jumps discussed by
VdKL99,00. Specifically, the
plateau seen here is associated with the balance between Fe~{\sc iii}
and Fe~{\sc ii}: 
in models on the low-density (i.e. low-$\Phi$) side of the plateau, 
Fe~{\sc iii} is the single most important ion for driving the wind while
on the high-density side Fe~{\sc ii} provides the largest contribution.
This transition in the driving
mechanism leads to the complex 
shape of the $\Lambda(\Phi)$ curve. 
 
These plateaus influence the effect of line branching on the derived 
mass-loss rates. Consider Models I and J: for these models, the vertical offset
between $\Lambda(\Phi)$ computed with and without branching is comparable
(see Fig. 4). For Model I, $\Lambda=1$ occurs at values of
$\Phi$ beyond the plateau, in a region where
$\Lambda(\Phi)$ varies rapidly with $\Phi$.
This leads to a relatively small difference
between $\Phi_{\mbox{\scriptsize res}}$ and $\Phi_{\mbox{\scriptsize bran}}$.
However, for Model J, $\Lambda=1$ lies in the region 
where $\Lambda(\Phi)$ varies very slowly. This means that the difference 
between $\Phi_{\mbox{\scriptsize res}}$ and $\Phi_{\mbox{\scriptsize bran}}$
is much greater for this model. Models K, L and M are all similar to, but
less extreme than, Model J. Conversely, Model N behaves similarly to 
Model I since it has $\Lambda=1$ significantly below the plateau.

Taken together, Models I -- N (and also the lower temperature Model O)
indicate that the effect of line branching is generally greater at
lower temperatures (B stars) than at high temperatures (O stars).
They also show that subtleties in the modelling (in particular
changes in the shape of  $\Lambda(\Phi)$ resulting 
from changes in the driving mechanism around,
for example, the bi-stability jumps) can influence the differential
effect of line branching such that it cannot be described by a simple scaling
with $\bar{\rho}$ and $T_{\mbox{\scriptsize eff}}$ alone.

For most of the models, the mass-loss rate computed with resonance
scattering agrees with that found by VdKL00 to within a factor of two or
better, which is adequate given the differential nature of the calculations
presented here. 
Those models where the agreement is slightly poorer
(Models L, K, N and M) 
all have self-consistent mass-loss rates
($\Phi_{\mbox{\scriptsize res}}$) which lie close to plateaus in
$\Lambda(\Phi)$ -- since $\Lambda(\Phi)$ varies slowly in these
regions relatively minor inadequacies in the models (i.e. the 
computation of $\Lambda$) lead to larger errors in the self-consistent
value of $\Phi$.
In any case, these errors are of little concern since it can
be seen from Fig. 4 that the differential effect of branching 
(the offset between the pairs of curves in the figure) is not a
strong function of $\Phi$.

In conclusion, for hot $\sim 40,000$~K stars with luminosity
$\sim 10^6 L_{\odot}$ calculations which neglect branching
overestimate the mass-loss rate by $\sim 20$ per cent while for 
lower luminosity stars with similar temperatures the effect is smaller.
The overestimation factor can be larger for the cooler
($\sim 20,000$~K) B stars with similar luminosities 
(typically about 50 per cent 
and in extreme cases up to a factor of a few) and can be significantly
affected by variations in the ionisation balance (and therefore
wind driving mechanism) in the models.

\section{Wolf-Rayet star}

W-R
stars have larger mass-loss rates (up to $\sim 10^{-4}$~M$_{\odot}$~yr$^{-1}$)
than OB-stars but comparable luminosities. LA93
and Gayley, Owocki \& Cranmer (1995) 
have
modelled line driving in W-R winds and shown that the high density means
that multiple scattering is very important in these winds (the wind
performance number is typically $\sim 7$, far in excess of the single
scattering limit). Thus, since the influence of line
branching is linked to the predominance of multiple scattering, W-R star
winds are likely to show a large differential effect if line
branching is introduced to the calculation. This effect is quantified here.

The W-R wind model adopted for this study is that of LA93
($M_*=12.6$~M$_{\odot}$, $R_{c}= 2.5$~R$_{\odot}$, 
$L_{*}=2.8 \times 10^5$~L$_{\odot}$,
$v_{c} = 10$~km~s$^{-1}$, $v_{\infty} = 2500$~km~s$^{-1}$, $\beta=1$).
These parameters were chosen by
LA93 to represent a generic WN5 star.
The model is similar to the OB-star models discussed above, the
main differences being the temperature stratification and the element
abundances. These are discussed briefly below.

As discussed by LA93 and subsequently confirmed by
Schulte-Ladbeck, Eenens \& Davis (1995) and
Herald et al. (2000), observations suggest that there is
a significant outward variation of ionisation in W-R winds. 
Ionisation stratification has also been predicted theoretically from W-R type
models (e.g. de Koter et al. 1997).
To account for 
this, LA93 imposed a simple variation of radiation temperature with
velocity (their equation 11) which is also used here. Their standard 
value for the radiation temperature at $v_{\infty}$ of
$\log T_{R}=4.35$ is adopted here. 
The outward variation of
electron temperature is less critical to the calculation: it is 
obtained from the modified Milne-Eddington temperature
distribution (Lucy 1976), following LA93.
In addition, the usual dilution factor ($W$) is replaced by the modified
dilution factor ($W_i$) defined by LA93.

The same elemental composition used by LA93 is adopted.
Thus $X=0.0$, $Y=0.98$ and $Z=0.02$. The carbon, nitrogen and oxygen
are taken from a WNE model (Maeder \& Meynet 1987) and the other metal 
abundances are scaled from solar values to give $Z=0.02$.

As in the OB-star models, the radiation field at the base of the model
$r=R_{c}$ is assumed to be black-body with a high energy cut-off at the
He~{\sc ii} ionisation edge. 
This cut-off is inconsistent with the adopted ionisation structure (i.e.
if there is really no radiation shortward of the He~{\sc ii} edge then there
are no photons to produce the high ionisation stages by
photoionisation).
However, this is of little consequence here since in reality there
will be a locally produced radiation field that can produce the high
ions.
In addition, the calculations presented here are relatively
insensitive to the wavelength cut-off in the incident radiation field
since most of the input energy is at longer wavelengths: this has been
checked explicitly with Monte Carlo simulations in which the high energy 
cut-off was placed at higher energy (by a factor of ten) -- the results of 
those simulations are not significantly different from those presented below.

The black-body temperature adopted for the 
incoming radiation field is $T_{\mbox{\scriptsize eff},c}$, the effective
temperature determined from $L_{*}$ at $R_{c}$. To allow for the higher
temperatures deep in W-R winds, the line list used for the OB stars (see
Section 4) was expanded to include lines from higher ionisation stages 
({\sc vii} and {\sc viii}) of the most important iron group elements 
(iron and nickel). In the calculations that follow,
however, these ion stages provide a negligible 
contribution ($< \sim 1$ per cent)
to the radiative driving because of the assumed
shape of the
input stellar radiation field -- most of the energy is 
at wavelengths that are too long for the packets to 
interact with the strong lines of these ions. 

Fig.~5 shows $\Lambda$ versus $\Phi$ for W-R models computed with and
without line branching in the Monte Carlo simulation. As expected, Fig.~5 
shows that line branching has a far larger effect on the mass-loss calculation
for a W-R star than for OB stars: the difference between
$\Lambda$ computed with and without branching for a given value of 
$\Phi$ is much greater than for any of the models discussed in Section 5.
The self-consistent W-R mass-loss rate 
determined without branching is $1.8 \times 10^{-5}$~$M_{\odot}$ yr$^{-1}$ 
(which is close to the value of $2.1 \times 10^{-5}$~$M_{\odot}$ yr$^{-1}$
found by LA93). When line branching is included the mass-loss 
rate is only $5.4 \times 10^{-6}$~$M_{\odot}$ yr$^{-1}$, a reduction by more 
than a factor of 3. Note that while some of the cool OB models discussed in
Section 5 also predicted large reduction factors
($\Phi_{\mbox{\scriptsize res}}/\Phi_{\mbox{\scriptsize bran}}$), 
these were not because branching had such a large effect on the computed
values of $\Lambda$ for a given value of $\Phi$ 
but were because of the relative insensitivity of $\Lambda$ to the value of
$\Phi$ adopted in the modelling.

\begin{figure}
\epsfig{file=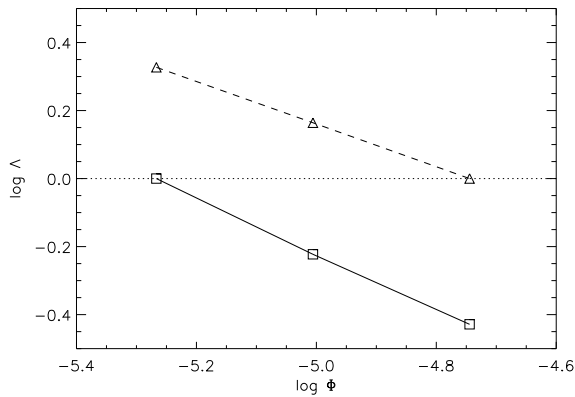,width=8cm}
\caption{$\Lambda$ as a function of $\Phi$ for the W-R model: 
the boxes (connected by solid
line) show computations with line branching and the triangles (connected 
with dashed line) show computations with resonance line scattering only.
The dotted horizontal line is $\Lambda=1.0$ (self-consistent model).
The $\Lambda$ values are accurate to around
$\pm 0.01$~dex.}
\end{figure}

The magnitude of this effect can be explained following the discussion of the
O stars above. Fig.~6 shows the distribution of the number of line 
scatterings for the W-R model. 
The reduction in the number of packets that are scattered many times as
a result of line branching is dramatic in the W-R model: without 
branching, individual packets undergo up to $\sim 700$ line scatterings (in a 
simulation with $10^6$ packets) but with branching included
no packets 
undergo more than 90 line scattering events.
Comparing Figs. 2 and 6, shows that
multiple scattering is much more important in the W-R model than in the
$\zeta$~Puppis model. Thus photon leakage from one part of the spectrum to
another has a much greater influence on the derived mass-loss
rate. 

\begin{figure}
\epsfig{file=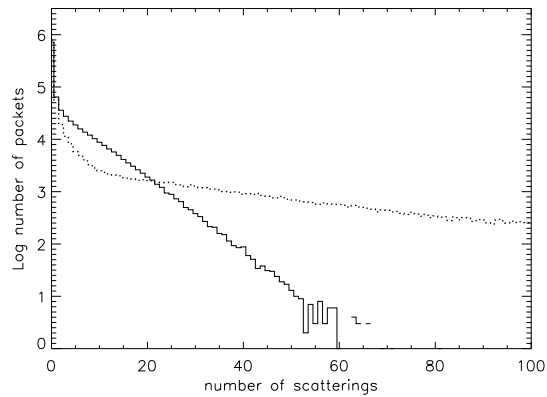,width=8cm}
\caption{The number of packets versus the number of line scatterings for 
the W-R model from Monte Carlo simulations with a total of $10^6$
packets. The solid histogram shows a simulation where line branching is 
allowed.
The dotted histogram shows the case where only resonance scattering is 
permitted.}
\end{figure}

\begin{figure}
\epsfig{file=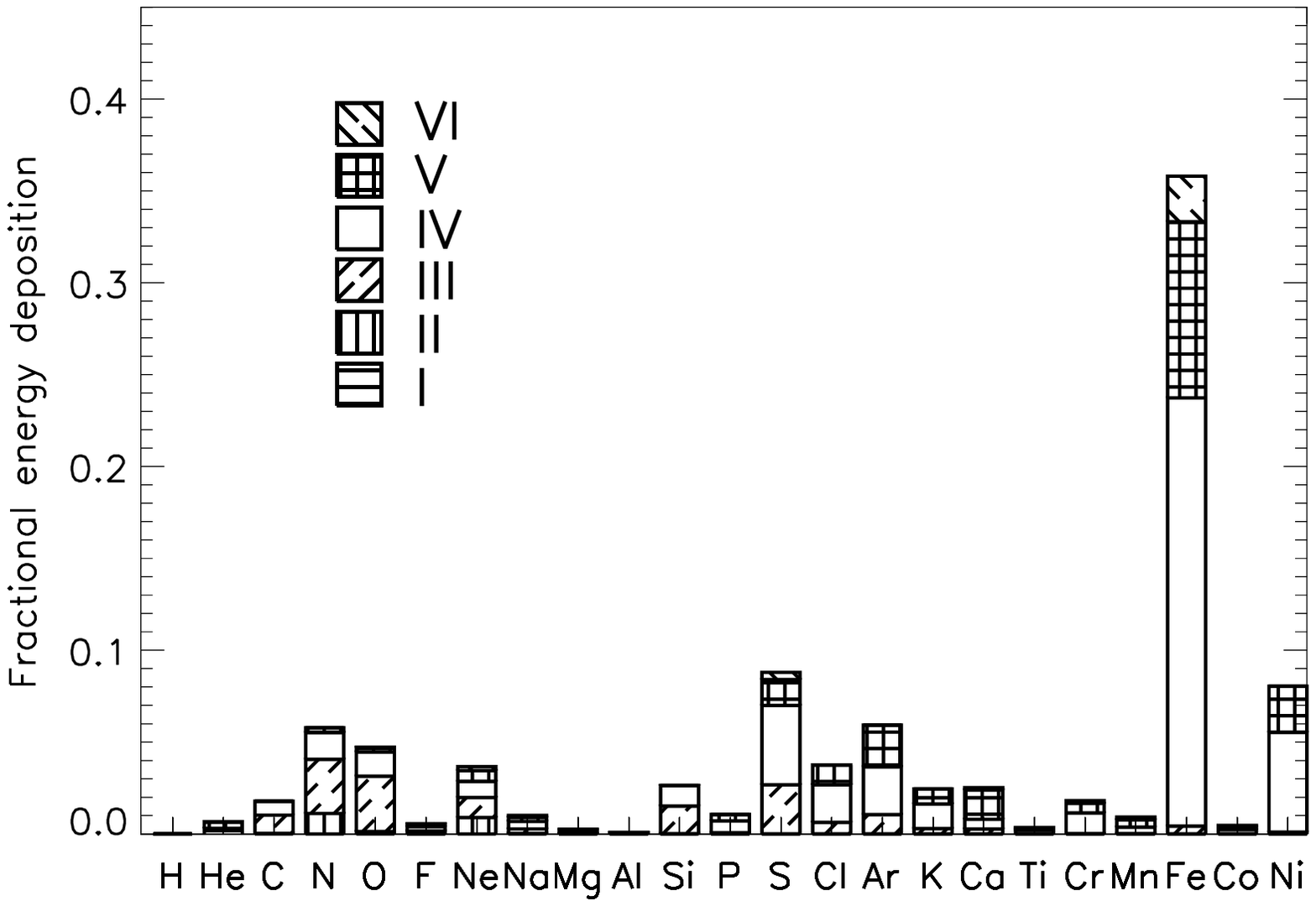,width=8cm}
\epsfig{file=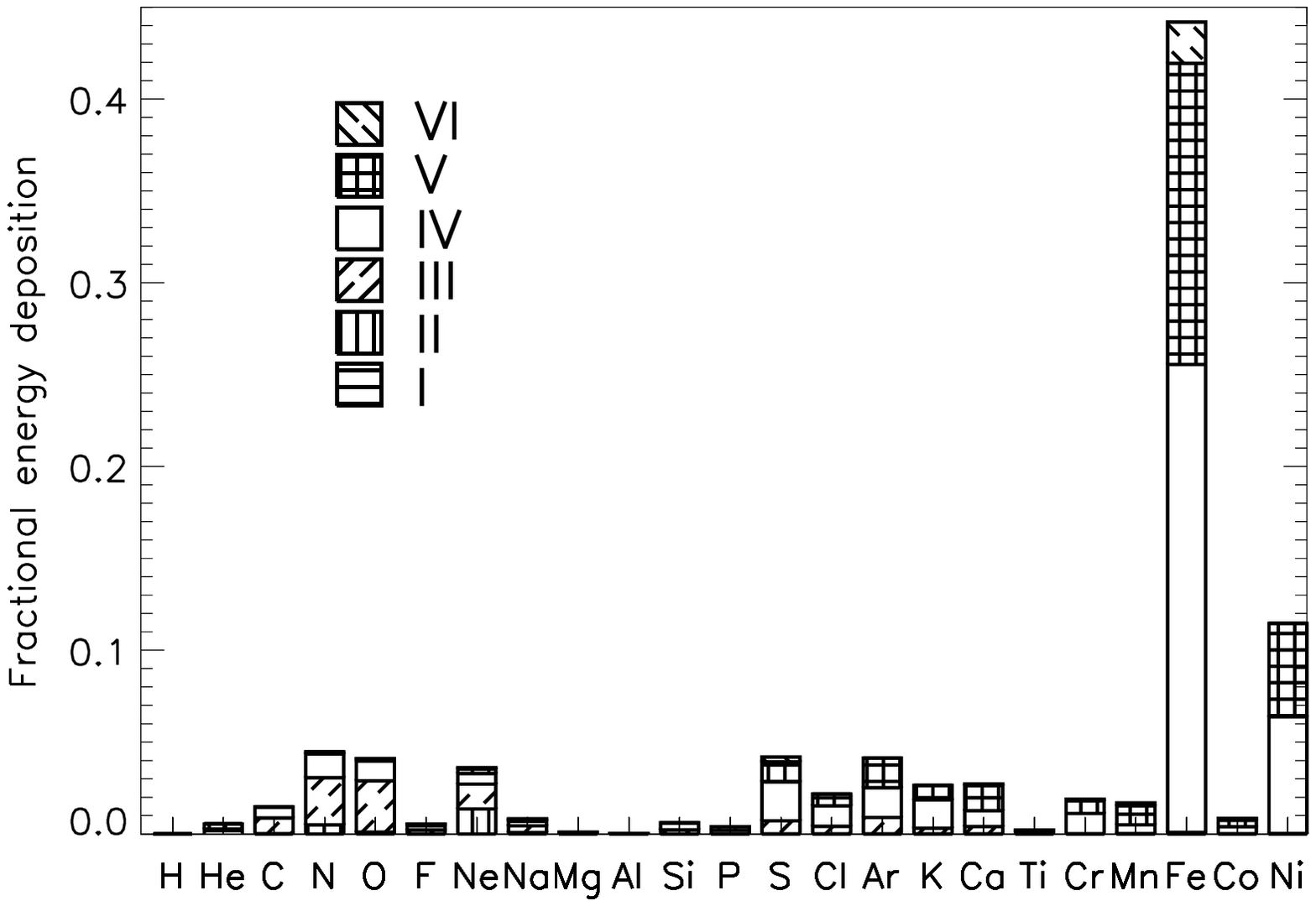,width=8cm}
\caption{The relative importance of the different elements/ions to the
mass-loss-rate for the W-R model. The upper panel shows the contributions
when branching is included and the lower panel shows the case where
only resonance scattering is included.
In both cases
the chart shows the fraction of energy deposited in the wind by each element.
The bars for each element are subdivided by ionisation stage. Note that 
although the seventh and eighth stages of iron and nickel
are included in the simulation they make
a negligible contribution which is not shown here. The fractions do not sum to 
1.0 because of the contribution of
scattering by free electrons (not shown).
}
\end{figure}

Fig.~7 shows the contribution of the different ions to the mass-loss rate 
for the W-R model. The lower panel is from a simulation
with resonant scattering only and the upper shows calculations with line
branching. The W-R mass-loss rate is dominated by the
iron group elements (Fe and Ni). When branching is included the
relative contribution of the iron group elements drops and the lighter
elements become more important: as discussed in Section 5.1, this is 
expected because of the greater complexity of the iron group ions. The effect 
here is more dramatic than in $\zeta$~Puppis, but is still small
(the fractional contribution of Fe and Ni to the line driving is
smaller when branching is allowed by only $\sim 10$ per cent).

There is some observational evidence to suggest that the wind acceleration
is more gradual in W-R stars than OB stars: e.g. based on infrared data
Ignace, Quigley \& Cassinelli (2003) have recently suggested $\beta$ values
of 2--3 for a WN6 star, and they do not rule out larger values. Therefore,
to investigate the sensitivity of the results presented here to the adopted 
value of $\beta$, calculations have been performed using the same W-R 
parameters as above except that $\beta=3$ was used for the velocity law.
In this case, the derived mass-loss rates were smaller: 
$7.2 \times 10^{-6}$~$M_{\odot}$ yr$^{-1}$ and 
$4.4 \times 10^{-5}$~$M_{\odot}$, with and without line branching, 
respectively. The differential effect of line branching is smaller
here than for the $\beta=1$ case, but is still significant 
(a factor of $\sim 1.6$).

\section{Conclusions}

In the previous sections, simple wind models for a range of hot stars have
been used to investigate the effect of line branching on mass-loss rates.
For O stars, where accurate mass-loss rates are important for 
evolutionary calculations, the overestimation is relatively small,
typically $<25$ per cent and is correlated with the predominance
of multiple scattering.
For
these stars the 
difference in the mass-loss rate computed with and without line branching 
is
comparable to or less than 
typical observational errors in O-star mass-loss rates. Thus,
despite the assumption of resonant scattering,
in most cases
the VdKL00 mass-loss calculations are not significantly in error.
In the future, however, 
when observational constraints become tighter it will be
necessary to include line branching in realistic mass-loss calculations.

The effect of line branching in 
relatively cool ($\sim 20,000$~K)  
luminous stars (early B stars) 
is significantly larger
than in the hotter stars (on average about 50 per cent), 
but it is also more sensitive to the details of the modelling.

Line branching is most important under conditions typical of a W-R star.
The W-R calculations presented here suggest that 
neglecting line branching 
results in an overestimation of the mass-loss rate by a 
factor $\sim 3$ for $\beta=1$. 
The effect is less severe if a larger value of $\beta$ is adopted but 
is still significant at least for $\beta \sim 3$.
This has implications for
modelling line driving in W-R stars: branching makes it significantly
harder for line driving to accelerate the winds of these stars. There are
many unanswered questions relating to W-R stars, 
particularly regarding  
the roles of line and continuum driving throughout their winds. But 
the results presented here clearly show that 
branching significantly influences mass-loss calculations which
involve multiple scattering and that future studies of line driving in 
W-R stars must incorporate these effects.

\section*{Acknowledgements}

I wish to thank L. Lucy for many useful discussions and insightful comments
relating to this work. Thanks also to J. Vink, J. Drew, S. Owocki and,
in particular, the referee A. de Koter for 
their comments. This work was carried out while I was a PPARC supported 
PDRA at Imperial College London (PPA/G/S/2000/00032).

\section*{Bibliography}

Abbott D. C., Lucy L. B., 1985, ApJ, 288, 679 (AL85)\\
Castor J. I., Abbott D. C., Klein R. I., 1975, ApJ, 195, 157\\
Crowther P. A., Dessart L., 1998, MNRAS, 296, 622\\
Crowther P. A., Pasquali A., De Marco O., Schmutz W.,\\
\indent Hillier D. J., de Koter A., 1999, A\&A, 350, 1007\\
de Koter A., Heap S. R., Hubeny I., 1997, ApJ, 477, 792\\
Gayley K. G., Owocki S. P., Cranmer S. R., 1995, ApJ, \\
\indent 442, 196\\
Herald J. E., Schulte-Ladbeck R. E., Eenens P. R. J.,\\
\indent Morris P., 2000, ApJS, 126, 469\\
Ignace R., Quigley M. F., Cassinelli J. P., 2003, ApJ, in\\
\indent press\\
Klein R. I., Castor J. I., 1978, ApJ, 220, 902\\
Kudritzki R.-P., Puls J., 2000, ARA\&A, 38, 613\\
Kurucz R. L., Bell B. 1995, Atomic Line Data CD-ROM\\
\indent No. 23, 
Cambridge, Mass.: S.A.O. \\
Lamers H. J. G. L. M., Leitherer C., 1993, ApJ, 412, 771\\
Lamers H. J. G. L. M., Nugis T., 2002, A\&A, 395, L1\\
Leitherer C. et al., 1996, PASP, 108, 996\\
Lucy L. B., 1976, ApJ, 205, 482\\ 
Lucy L. B., Solomon P. M., 1970, ApJ, 159, 879\\
Lucy L. B., Abbott D. C., 1993, ApJ, 405, 748 (LA93)\\
Lucy L. B., 1999a, A\&A, 344, 282\\
Lucy L. B., 1999b, A\&A, 345, 211\\
Lucy L. B., 2002, A\&A, 384, 725\\
Lucy L. B., 2003, A\&A, 403, 261\\
Maeder A., Meynet G., 1987, A\&A, 182, 243\\
Morton D. C., 1967, ApJ, 147, 1017\\
Nugis T., Lamers H. J. G. L., 2002, A\&A, 389, 162\\
Onifer A. J., Gayley K. G., 2003, ApJ, 590, 473\\
Owocki S. P., Gayley K. G., 1999, in IAU Symp. 193:\\
\indent `Wolf-Rayet Phenomena in 
Massive Stars and\\
\indent Starburst Galaxies', p. 157\\
Puls J. et al., 1996, A\&A, 305, 171\\
Schaerer D., de Koter A., 1997, A\&A, 322, 598\\
Schulte-Ladbeck R. E., Eenens P. R. J., Davis K., 1995,\\
\indent ApJ,454, 917\\
Springmann U., Puls J., 1998, ASP conf. Ser. 131, 286\\
Vink J. S., de Koter A., Lamers H. J. G. L. M., 1999, A\&A, \\
\indent 350, 181 (VdKL99)\\
Vink J. S., de Koter A., Lamers H. J. G. L. M., 2000, A\&A, \\
\indent 362, 295 (VdKL00)\\
Vink J. S., de Koter A., Lamers H. J. G. L. M., 2001, A\&A, \\
\indent 369, 574\\
Vink J. S., de Koter A., 2002, A\&A, 393, 543\\

\label{lastpage}
\end{document}